\documentclass[prb,preprint]{revtex4-1} 

\usepackage[english]{babel}
\usepackage[utf8]{inputenc}
\usepackage{graphicx}
\usepackage{subcaption}
\usepackage{caption}
\usepackage{amsmath}  
\usepackage{amsfonts} 
\usepackage{amssymb}
\usepackage{xcolor,soul}
\usepackage{physics}

\begin{document}

\title{On Studies of Entropy of Classical and Quantum Kac Rings}
\author{Niamat Gill}
\affiliation{Amity International School, Noida, INDIA}
\author{Nishchal Dwivedi}
\email{nishchal.dwivedi@nmims.edu}
\affiliation{Department of Basic Science and Humanities, SVKM’s NMIMS Mukesh Patel School of Technology Management \& Engineering, Mumbai, INDIA }

\date{}

\begin{abstract}
Statistical physics is important in understanding the physics of interacting many bodies. This has been historically developed by attempts to understand colliding gases and quantifying quantities like entropy, free energy, and other thermodynamic quantities. An important contribution in statistical physics was by Boltzmann in the form of the H-theorem, which considered collisions between particles and used the assumption of molecular chaos or \textit{Stosszahlansatz} to understand macroscopic irreversibility. To elucidate these ideas, Mark Kac introduced a classical analog called Kac rings.\\

In this work, we attempt to introduce \textit{quantum-ness} in a Kac ring and study its entropy and recurrence, comparing and contrasting to corresponding trends in a classical Kac ring. We look at the trends of recurrence time for a system with a qubit as a pointer. We further study the time distribution of entropy for these systems.
\end{abstract}

\maketitle
\section{Introduction}
Atoms and molecules are the building blocks of the universe. These particles have microscopic properties which interact with each other in a complex, many-body fashion. Such a many-body system exhibit chaos, and thus, predicting their long-time futures by factoring every particle's equation of motion is difficult. To get some insight of such systems, we look at statistical quantities like temperature, pressure, entropy, etc., which give a good understanding of the behavior of the collective properties of these systems of particles. Although these statistical pictures are useful, witnessing and understanding the contribution of microscopic properties to this macroscopic nature is very interesting for researchers.\\
\\
One such model used to develop an intuition about the evolution of systems was given by Mark Kac \cite{kac1956some} as a classical analog to understand Boltzmann's H- Theorem \cite{kac1959probability, jebeile2020kac} and statistical physics \cite{gottwald2009boltzmann,chandrasekaran2012kac}. This model is known as a Kac Ring. \\
\\
A Kac ring is realized as a set of sites arranged on a rotating circular ring. Each site is occupied by a ball of one of two colors, say black and white, forming a one-dimensional periodic lattice. One of these sites is marked by a pointer. \\
\\
As the system evolves with a discrete clock, the ring rotates one ball at a time during every timestep in a single direction of rotation. When a ball passes the site with the pointer, it switches color. If there are $N$ balls on the ring, then after $2N$ time steps, the ring will return to its original configuration. This recurrence of the state of the ring implies a similar periodicity in thermodynamic properties, especially entropy, essentially demonstrating a case of Poincar\'e recurrence\cite{poincare1890probleme}.\\
\begin{figure}[h!]
  \centering
  \begin{subfigure}[b]{0.2\textwidth}
    \includegraphics[width=\textwidth]{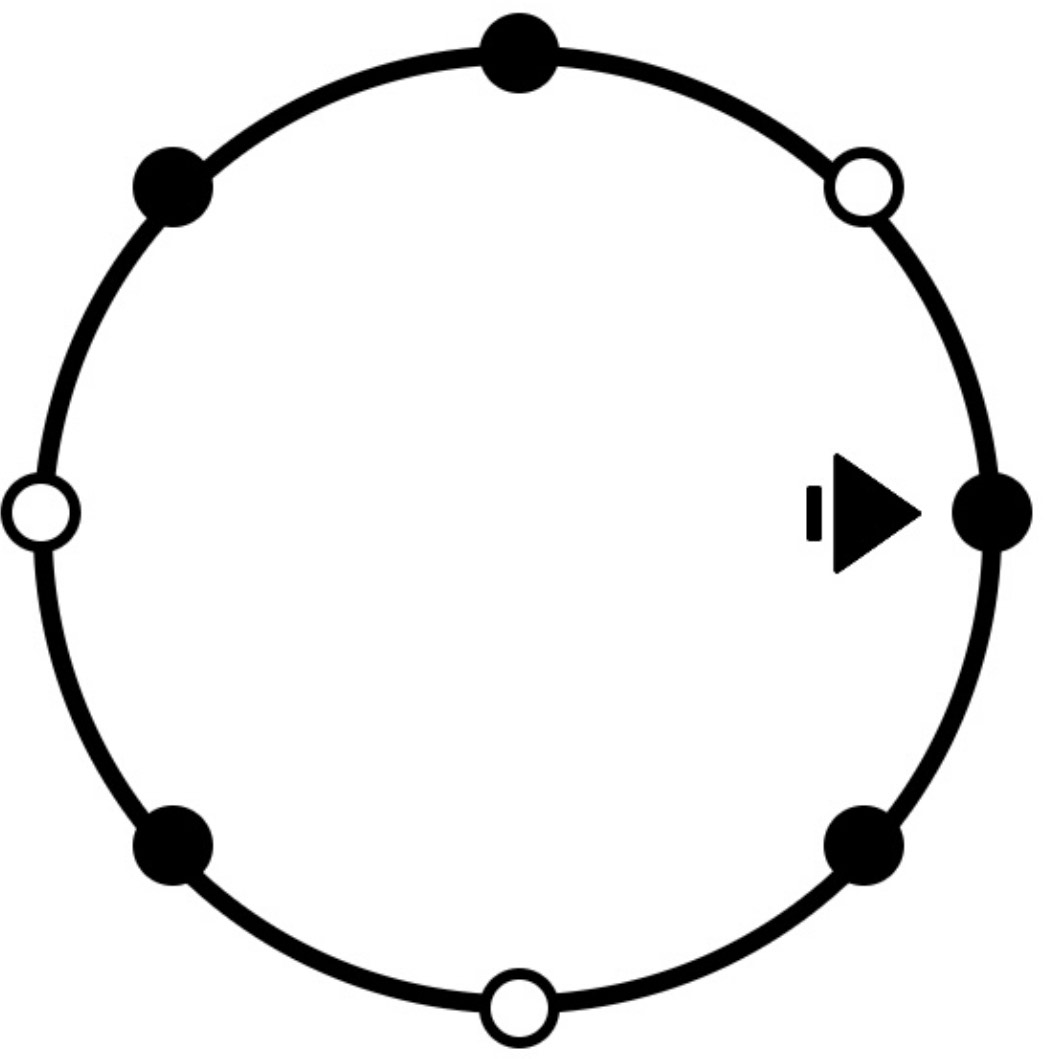}
    \caption{Initial State of Ring}
    \label{ring1}
  \end{subfigure}%
  \hfill
  \begin{subfigure}[b]{0.2\textwidth}
    \includegraphics[width=\textwidth]{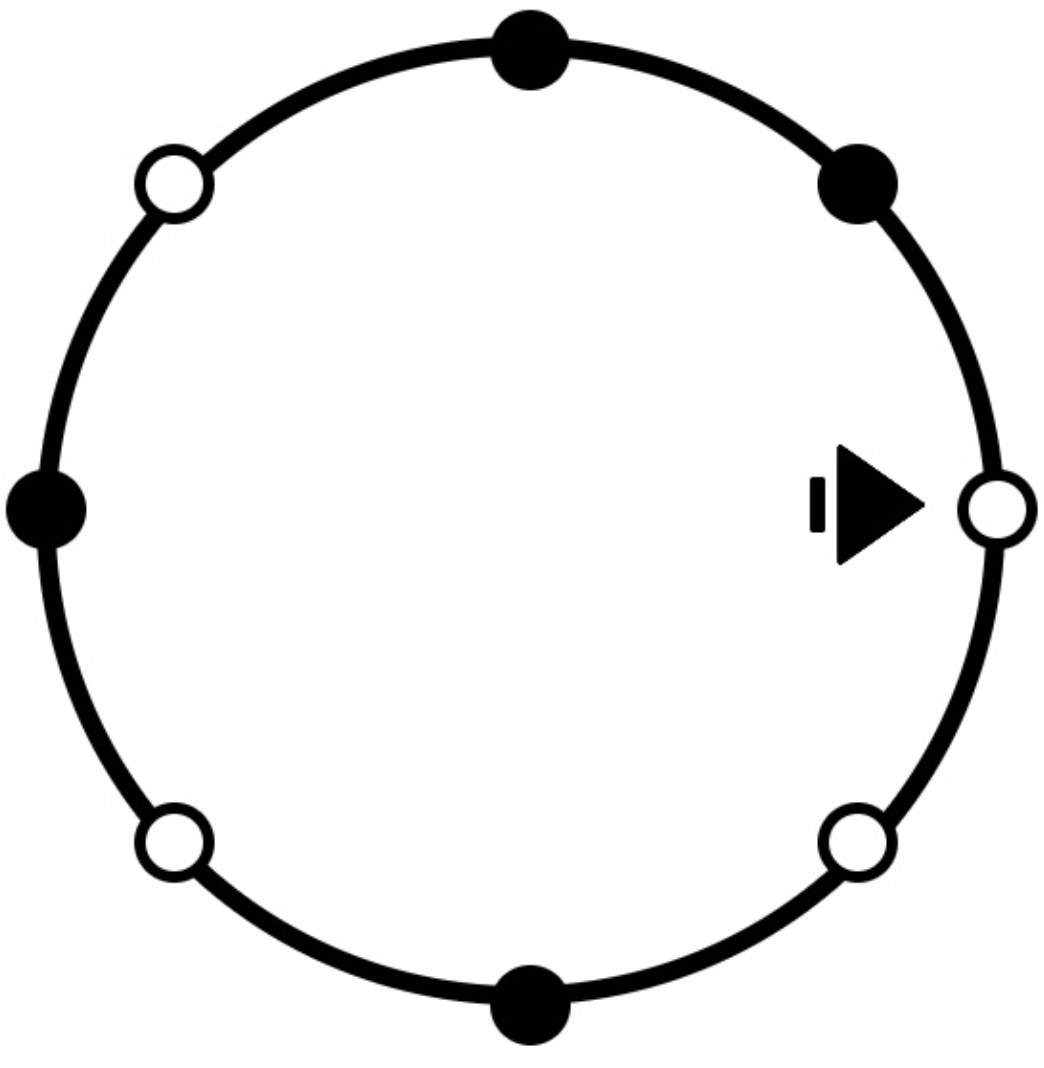}
    \caption{After 1 Timestep}
    \label{ring2}
  \end{subfigure}%
  \hfill
  \begin{subfigure}[b]{0.2\textwidth}
    \includegraphics[width=\textwidth]{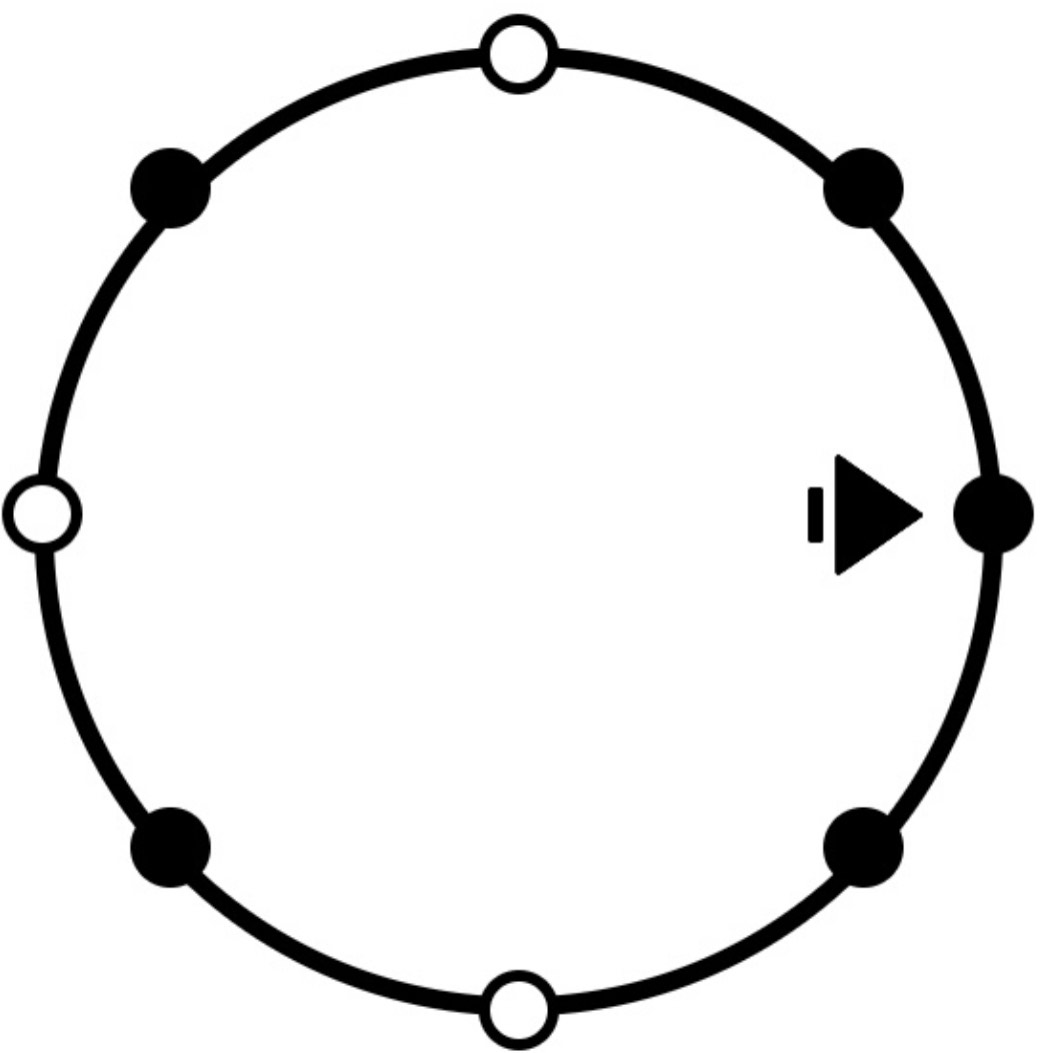}
    \caption{After 2 time steps}
    \label{ring3}
  \end{subfigure}%
  \hfill
  \begin{subfigure}[b]{0.2\textwidth}
    \includegraphics[width=\textwidth]{finalfinal/2.jpeg}
    \caption{After $2N$ time steps}
    \label{ring4}
  \end{subfigure}
  \caption{A Kac ring rotating in clockwise direction. After each time step, the balls rotate by site in the clockwise direction. After crossing the pointer, the ball changes its color. After $2N$ time steps, the original configuration is obtained again.}
  \label{rotating}
\end{figure}

Such Kac rings have been modeled extensively\cite{gottwald2009boltzmann}. Consider one such Kac ring with `$a$' balls and `$p$' pointers. Let $B(t)$ represent the number of black balls on a ring at time $t$ and $W(t)$ be a similar function for white balls. Let $b(t)$ be the number of black balls in front of a pointer at time $t$ (either 0 or 1 in case of a single pointer) and $w(t)$ represent the same for white balls. Then, the evolution of the ring can be modeled as:

\begin{eqnarray}
B(t + 1) = B(t) + w(t) - b(t) \\
W(t + 1) = W(t) + b(t) - w(t)
\end{eqnarray}
    
Following \cite{gottwald2009boltzmann} we define the quantity $\Delta(t)$ as difference between $B(t)$ and $W(t)$:
\begin{eqnarray}
\Delta(t) &=& B(t)- W(t)\\
\therefore \Delta(t+1) &=& B(t+1)- W(t+1)\\
\therefore \Delta(t+1) &=& \Delta(t) + 2w(t) - 2b(t)
\end{eqnarray}
Let $\mu$ indicate the probability of a site being marked by a pointer. 
\begin{eqnarray}
\mu &=& \frac{1}{N} = \frac{b}{B} = \frac{w}{W}\\
\therefore \Delta(t+1) &=& \Delta(t) + 2\mu W(t) - 2\mu B(t)\\
\therefore \Delta(t+1) &=& (1 - 2 \mu)\Delta(t)
\end{eqnarray}

Using the above relation, we can write $\Delta(t)$ as a recursive relation with $\Delta(0)$ as 
\begin{equation}
    \Delta(t)=\Delta(0) (1-2\mu)^t
\end{equation}
This implies that as the system evolves, $t\to\infty$, $\Delta(t)$ approaches 0 and the system must eventually have an equal number of black and white balls. In seeming violation of this conclusion, Kac rings demonstrate recurrence and return to their original configuration. The mathematics of deriving the time period of recurrence is beautifully derived in \cite{gottwald2009boltzmann}.\\
\\
Note that such a ring can be constructed with multiple colors \cite{jain2017kac} and multiple pointers \cite{gottwald2009boltzmann}. In this study, however, we only consider rings with two colors and a single pointer.\\
\\
Though a Kac Ring is a toy model, it is an accurate tool \cite{jebeile2020kac}, at least in the context of Boltzmann's H-Theorem, which is itself applicable to ideal gas systems initially at low entropy. Thus, Kac Rings can be useful in examining the evolution of systems of Quantum particles as well, provided appropriate modifications are made. 

\section{Method}
We consider a Kac ring with a randomized distribution of black and white balls. Each random initial distribution is taken as an element of an \textit{ensemble} and the results study the statistics of such an Ensemble of Kac rings.  \\
The classical Kac ring, with $N$ sites occupied by $N$ balls, rotates and a single ball passes over the pointer at each time step and switches its color- a white ball to black and a black one to white.
\\
Our \textit{Quantum} Kac ring follows a similar prescription, except that the pointer exists in a quantum superposition. 
%\hl{A few studies have discussed Quantum Kac rings where the balls also exist in a state of superposition }\cite{oberreuter2014entanglement}. 

We consider a state $\ket{0}$. A Hadamard gate is applied on this state to make a superposition of equal probability and this becomes the driving state of the pointer. We measure this superposition at every timestep. If the state collapses to $\ket{1}$, the ball at the pointer switches colors and if the state collapses to $\ket{0}$, the color of the ball remains the same.  After each time step, the state is reset post measurement. This system can be thought of a series of atoms of a similar state coming out of an oven, undergoing a superposition due to Hadamard gate. Each atom is measured one at a time, and they collapse to a state of $\ket{0}$ or $\ket{1}$. The measurement outcome of these atoms dictates if the state of the Kac ring changes in that time step or not. This toy model can be thought of as an interaction of a quantum state, where different quantum attributes (say, spin) interact differently with the system to give different outcomes. We study the statistics of such a toy model.
\\

%We initially wrote a code using Python to simulate Classical Kac Rings and after some preliminary exploration, used IBM's QISKIT module to simulate Quantum Kac Rings. QISKIT provides a set of tools and libraries for working with quantum circuits, including creating, simulating, and executing quantum circuits on actual quantum devices.\\
IBM's QISKIT \cite{qiskit} and various Python modules were used to run  simulations. The initial configuration was randomly generated with a random distribution of black and white balls. 
%The initial, randomly generated, configuration is stored in a list. Similarly, the current state of the ring- the color of balls in the sites- is stored in another list and compared to the original configuration after every timestep. A few conditional statements are used to change the color of the ball passing over the pointer. Since the ring is supposed to rotate as each ball passes over the pointer, the corresponding information stored in the list is also swapped in a manner such that it accurately represents the state of the ring.\\
We define a relative entropy as the relative change from the original configuration. In this way, whenever the original state is revisited, the entropy is zero.
This is done by counting the number of sites where the current color of the ball is different from that of the ball in the site originally. For example, in figure \ref{rotating} (b), we notice that the balls in 7 sites are of a different color as compared to the original configuration (a). Thus, after a single timestep, this particular configuration has a relative entropy of 7. In (c), 2 sites have a ball of a different color than (a), and thus a relative entropy value of 2 as compared to the initial configuration of the ring. \\
\\
Such a relative entropy is additive (if there are two rings in the original state and they are compared with their time evolved version, the total number of dissimilar balls for the system of two rings will be the same as the sum of dissimilar balls of each of them considered individually), scales with the increase in the number of balls, and depends solely on the configuration of the system.\\
\\
As the system evolves, the time steps for recurrence and relative entropy of the system evolves. We measured a large number of runs for an ensemble of such systems.

\section{Results and Discussion}

\subsection{time steps for Poincar\'e Recurrence}
When the state of the Kac ring returns to the initial configuration, the relative entropy  returns to its initial value. This is Poincar\'e recurrence. We study the time steps taken for this recurrence.

\subsubsection{Classical Kac Ring}

\begin{figure}[h!]
\centering
\includegraphics[width=0.8\columnwidth]{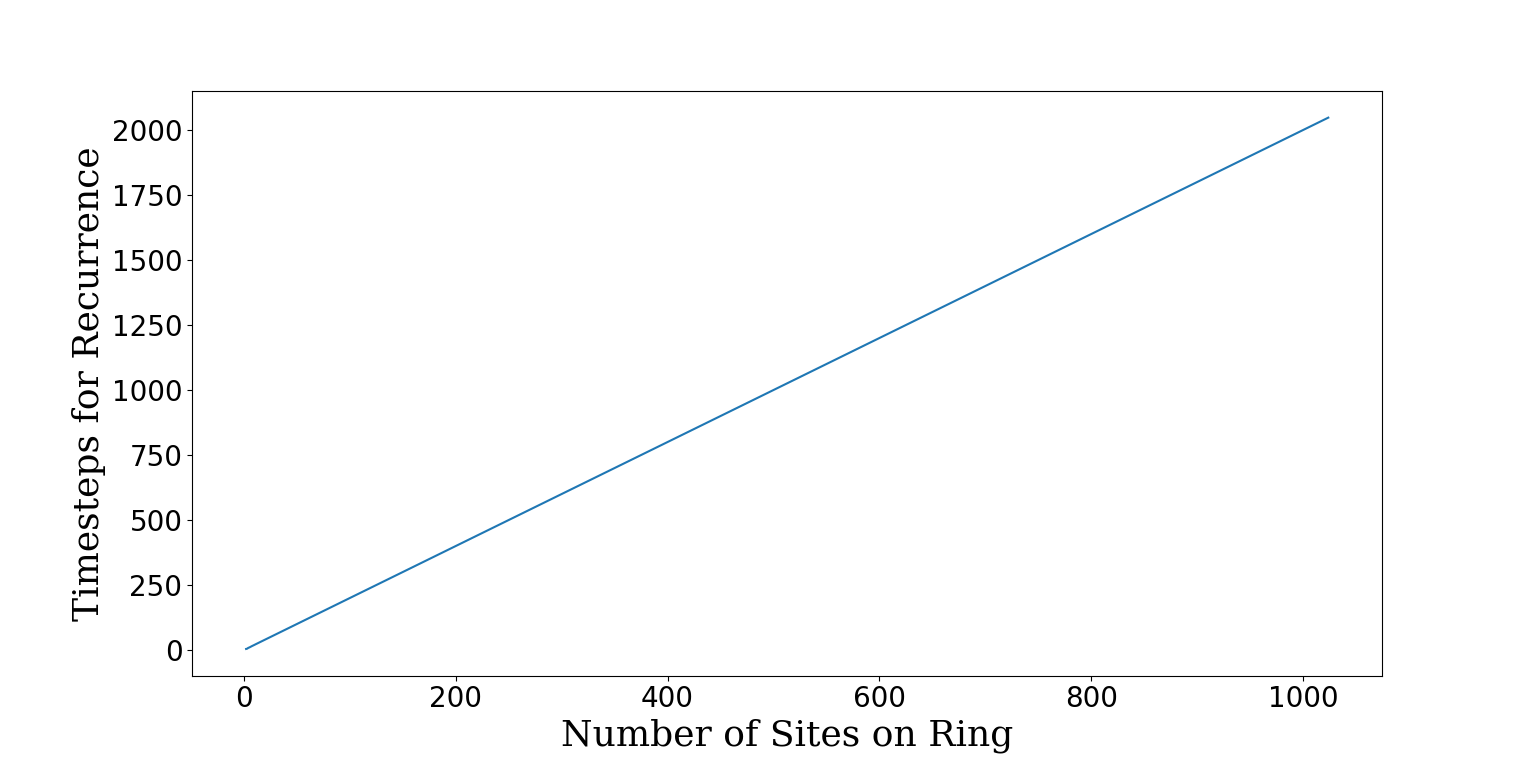}
\captionsetup{justification=centering} % center the caption
\caption{time steps for Poincar\'e recurrence as the number of sites increase for a classical Kac ring. Each site has a single ball. The relation is a linear graph of slope 2.}
\label{s1cl}
\end{figure}

As discussed before, \cite{gottwald2009boltzmann} a Classical Kac ring with $N$ balls eventually returns to its original configuration after $2N$ time steps or 2 complete rotations. We observe a similar trend even as $N$ becomes extremely large (figure \ref{s1cl}). The relation of the recurrence time to the number of balls on the ring is a simple linear relation.

\subsubsection{Quantum Kac Ring}

\begin{figure}[h!]
\centering
\includegraphics[width=0.8\columnwidth]{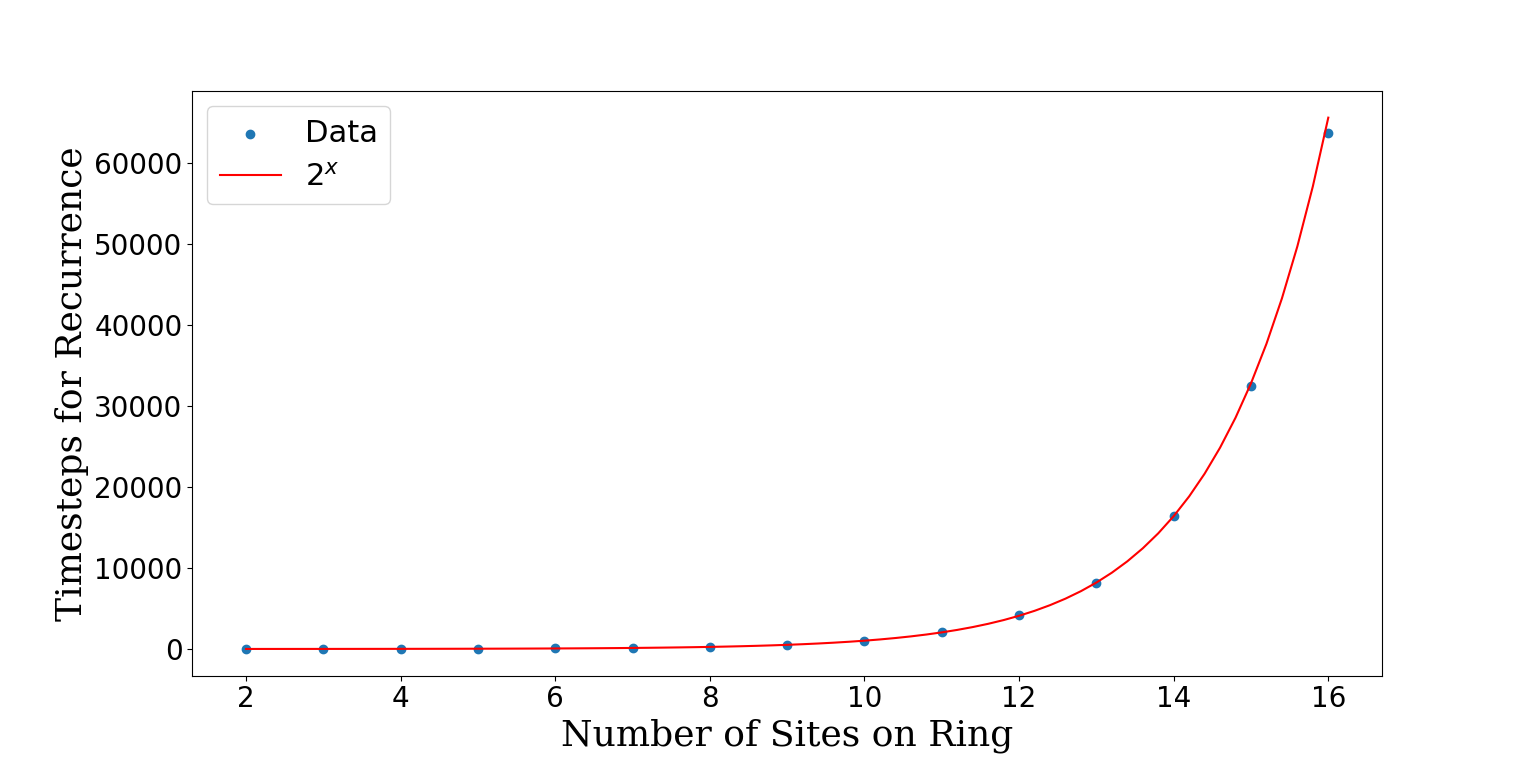}
\captionsetup{justification=centering} % center the caption
\caption{time steps for Poincar\'e Recurrence as the number of sites (each having a single ball) increases for a Quantum Kac ring. The quantum simulations fit perfectly to $2^x$, $x$ being the number of balls.} 
\label{s1qt}
\end{figure}

Just as in the case of the Classical Kac Ring, the relative entropy returning to 0 in Quantum Kac Ring will demonstrate Poincar\'e recurrence. Indeed, we observe that Quantum Kac ring, too, return to their original configuration. However, unlike in the case of classical rings, the recurrence time for an ensemble of runs as the number of sites on the rings increases follows a power law ($2^N$) rather than a linear one (figure \ref{s1qt}).

\subsection{Distribution of recurrence time}

We collect many instances of different initial states. Each of these states are evolved with a classical and a quantum pointer. We run this instance till the original configuration is revisited by the Kac ring. The time steps taken for this recurrence are collected and studied.

\subsubsection{Classical Kac Ring}

\begin{figure}[h!]
\centering
\begin{tabular}{cc}
\includegraphics[width=0.4\textwidth]{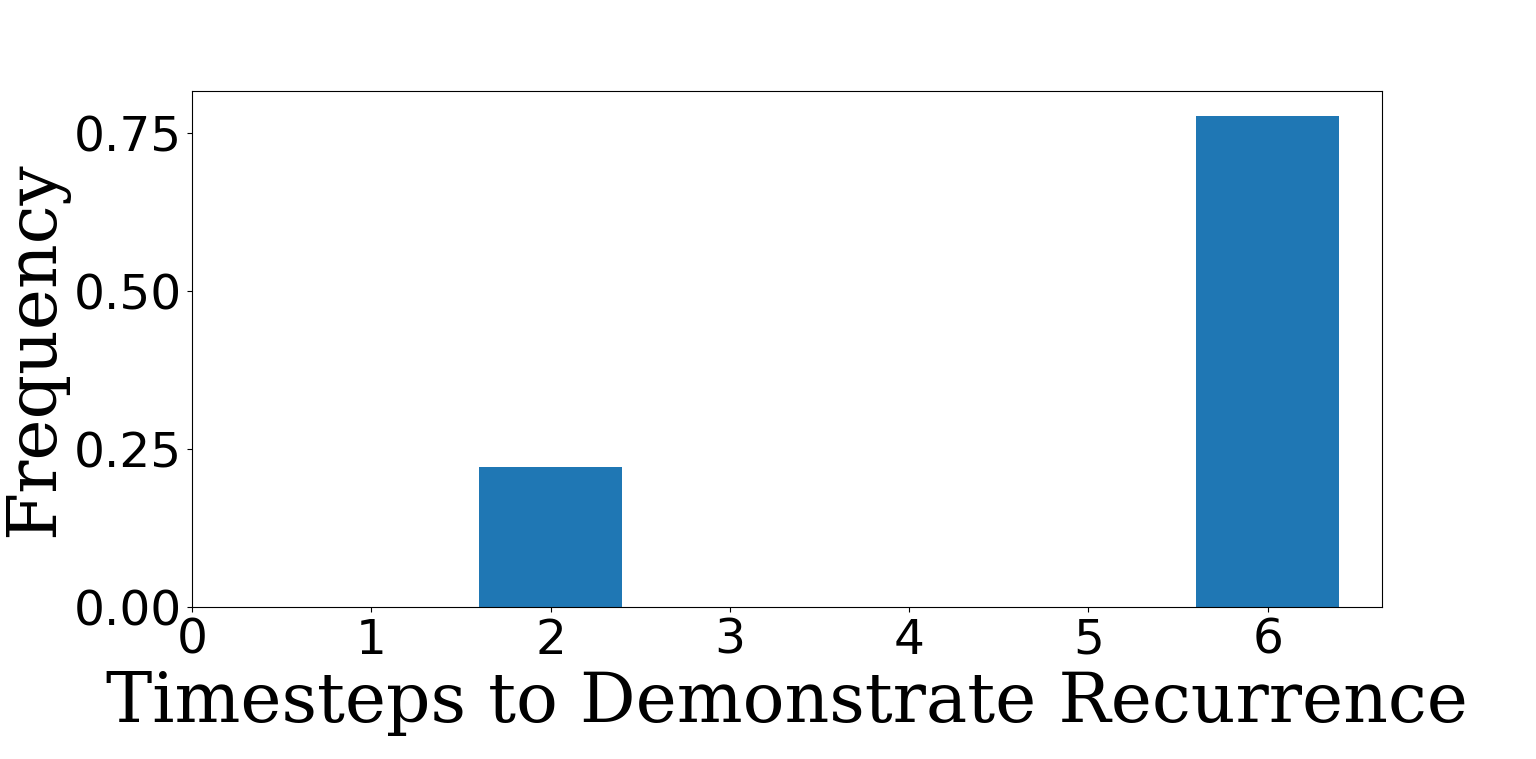} &
\includegraphics[width=0.4\textwidth]{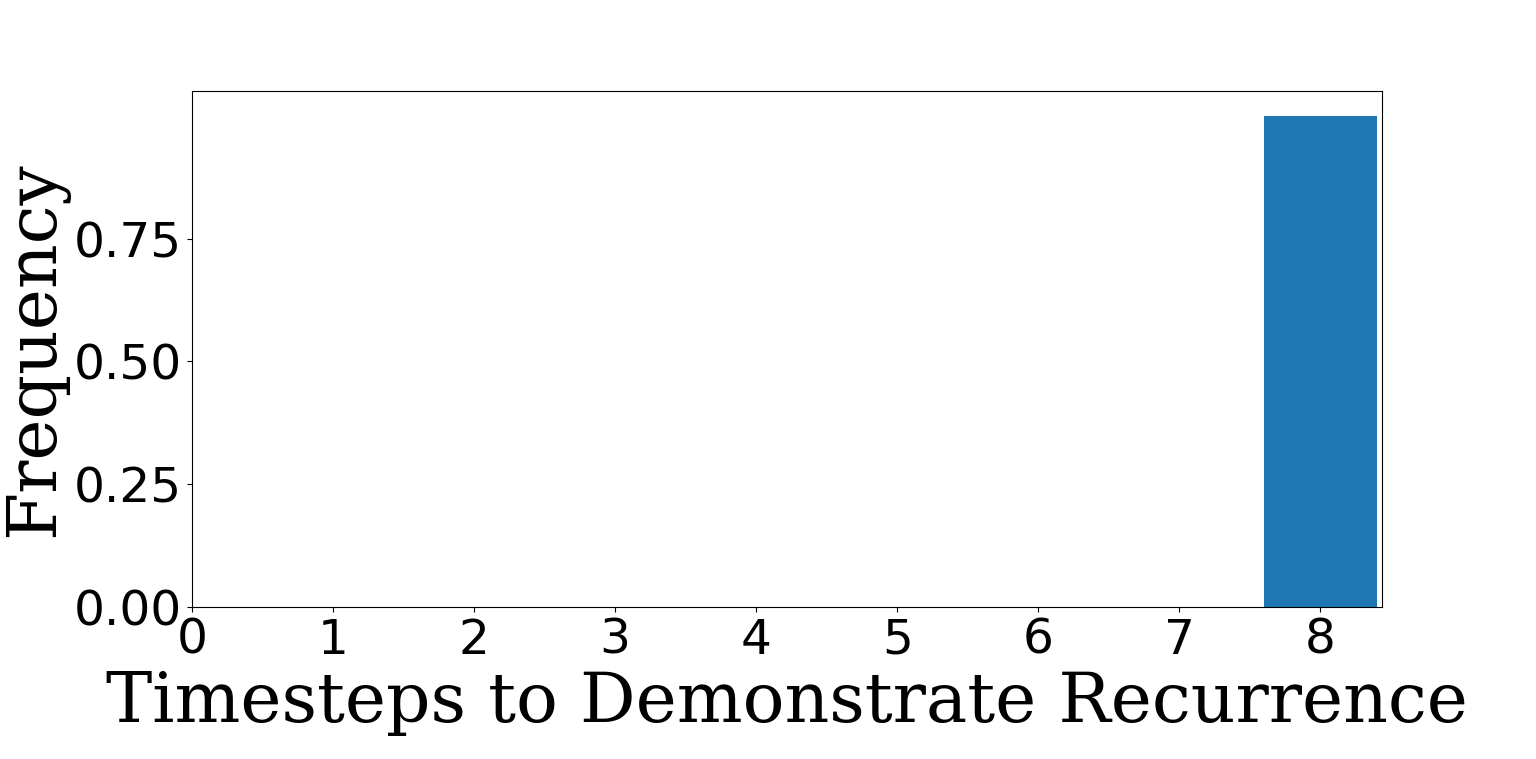} \\
(a) Classical Ring with 3 Sites & (b) Classical Ring with 4 Sites \\
\includegraphics[width=0.4\textwidth]{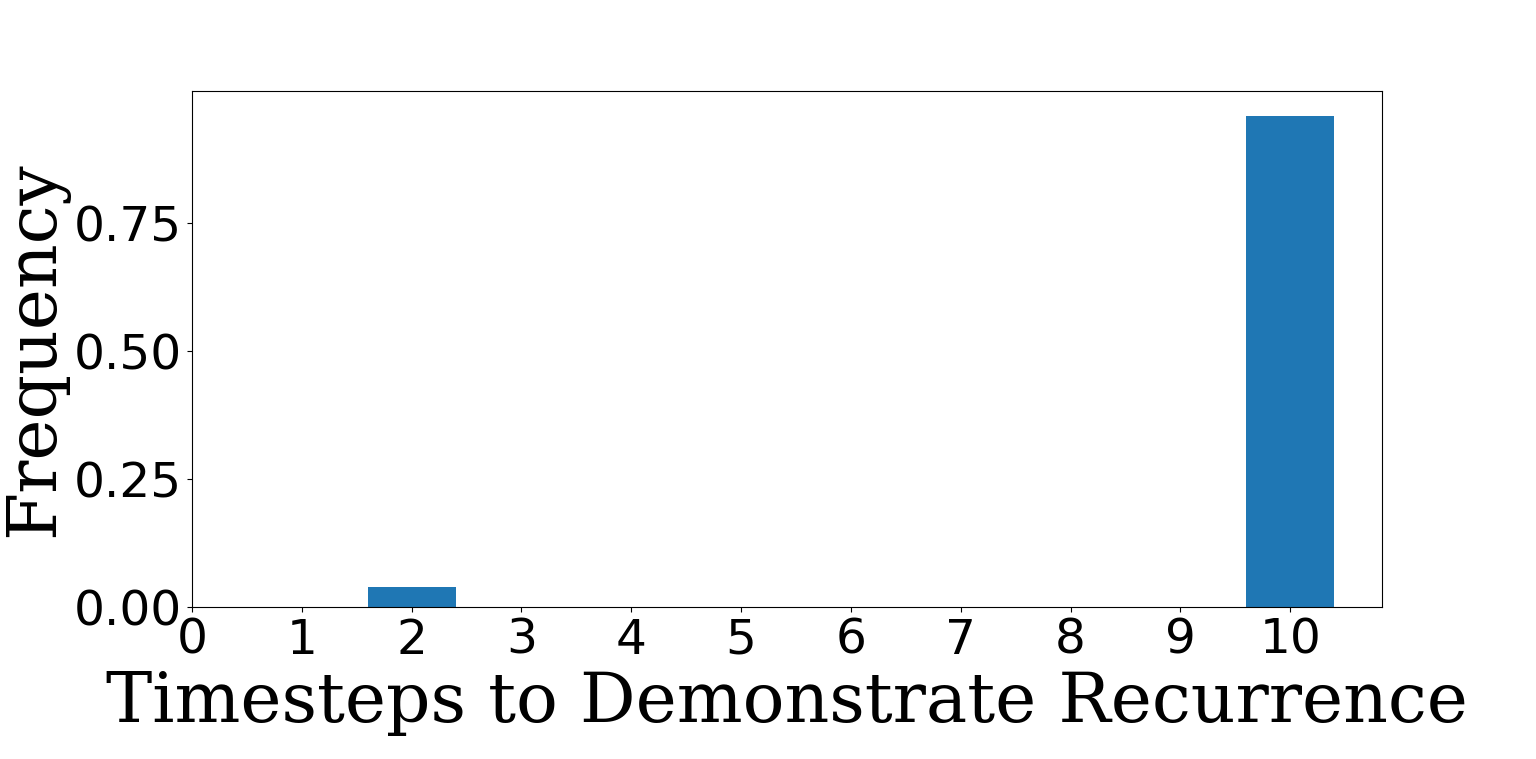} &
\includegraphics[width=0.4\textwidth]{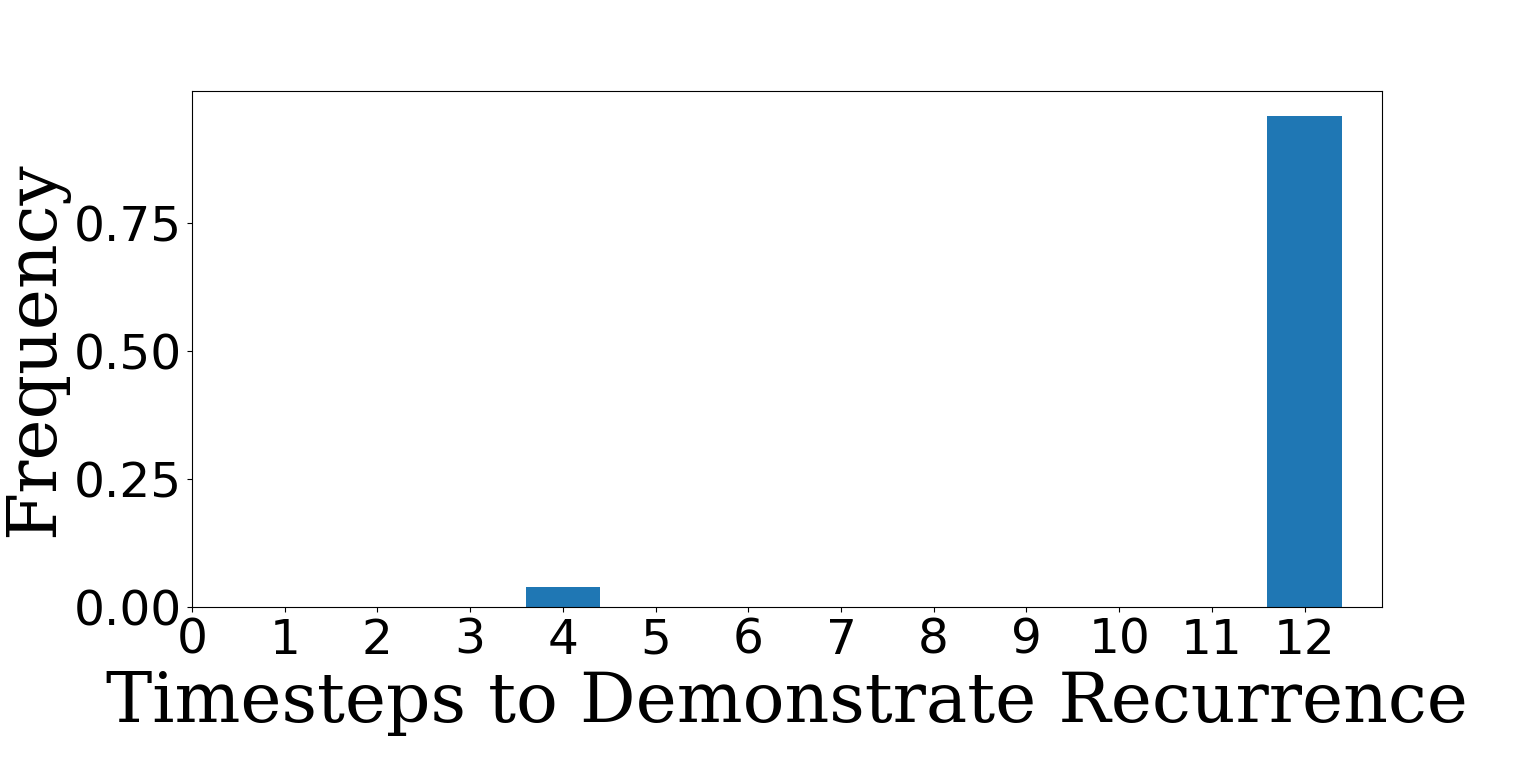} \\
(c) Classical Ring with 5 Sites & (d) Classical Ring with 6 Sites \\
\includegraphics[width=0.4\textwidth]{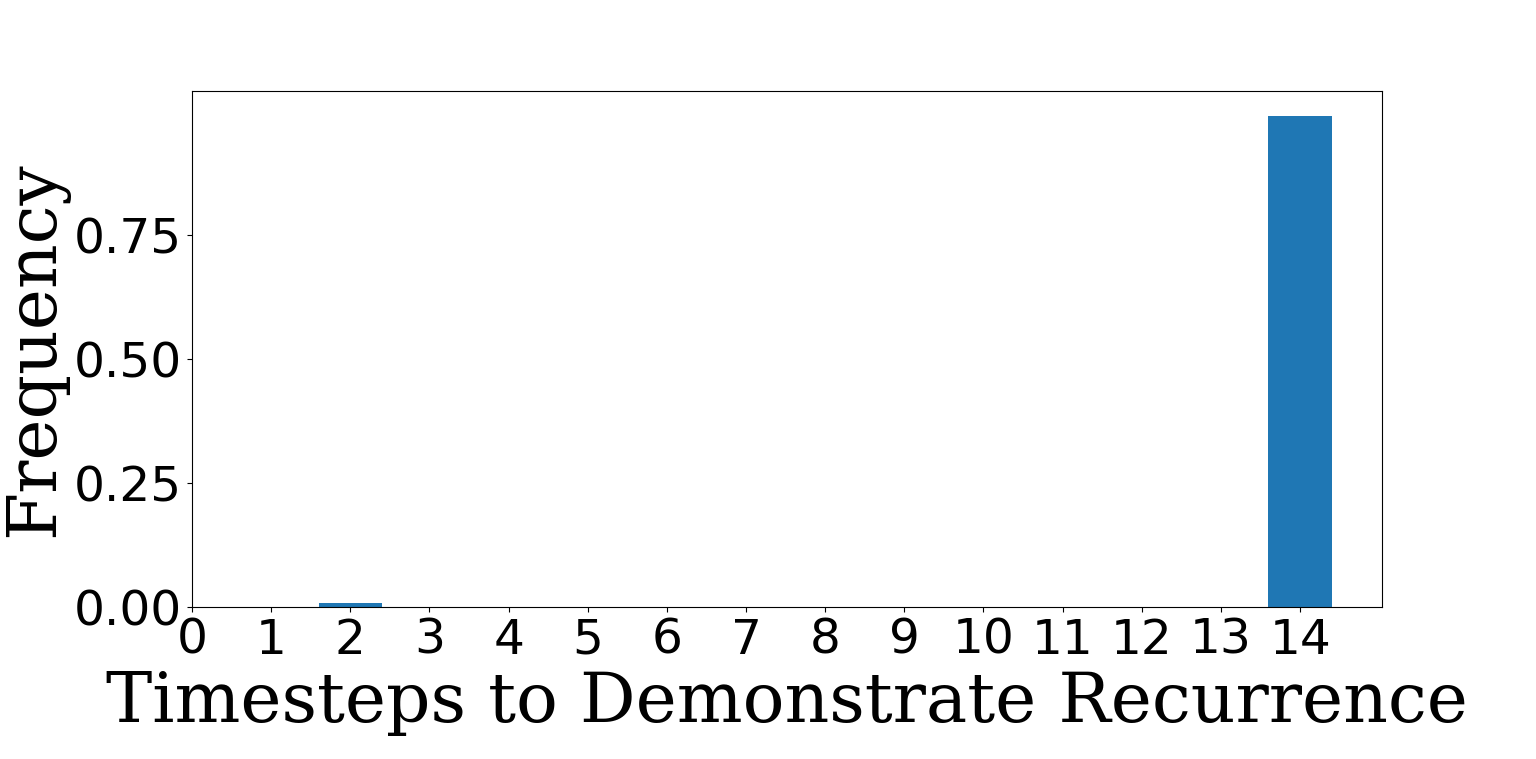} &
\includegraphics[width=0.4\textwidth]{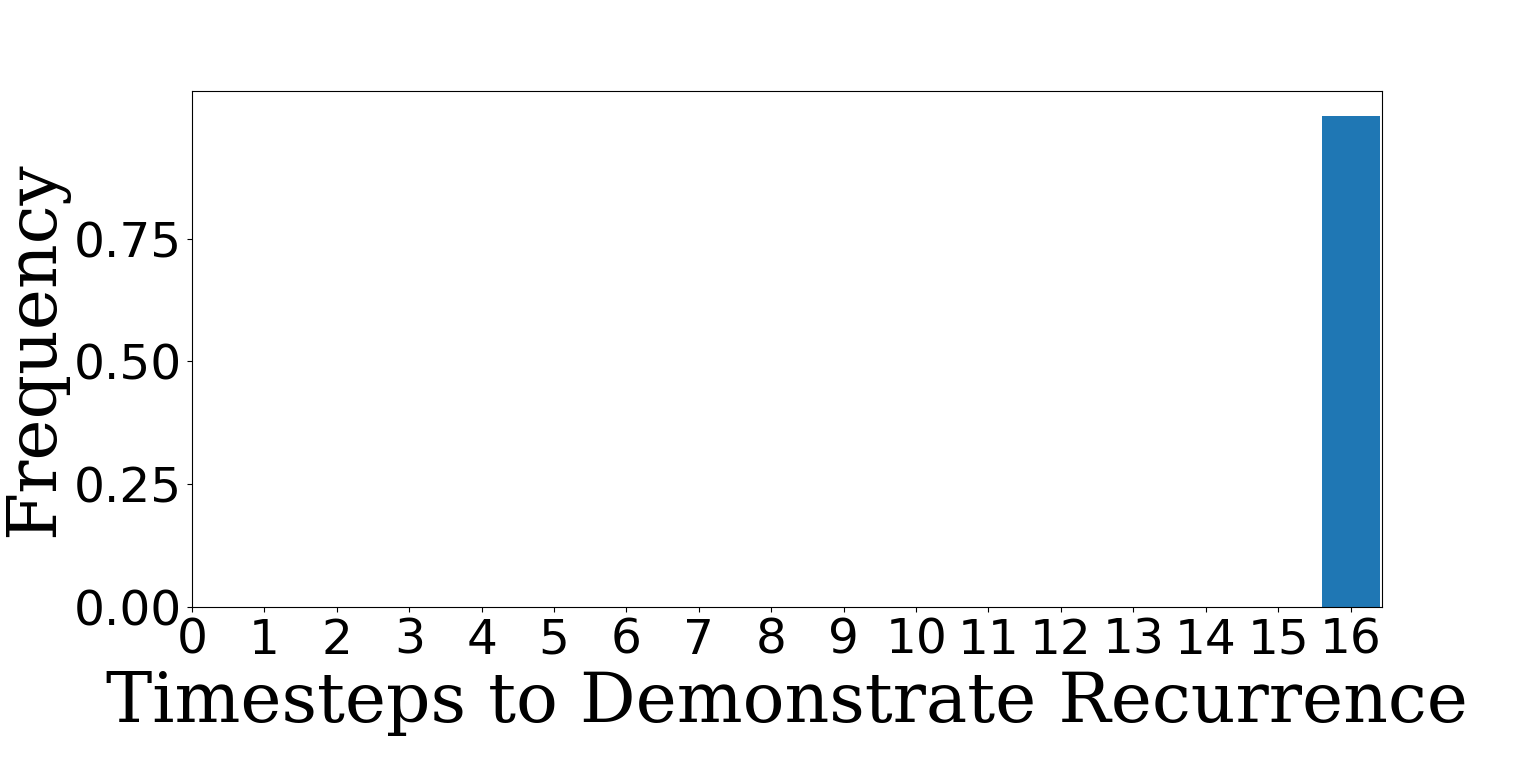} \\
(e) Classical Ring with 7 Sites & (f) Classical Ring with 8 Sites \\
\end{tabular}
\caption{Distribution of time steps for Poincar\'e Recurrence for a Classical Ring}
\label{s2cl}
\end{figure}

While most initial configurations show recurrence after $2N$ time steps, some do sooner. This happens purely because of the structure of the initial configuration, which can match the initial configuration after just a few steps of color change by the pointers. However, as $N$ increases, the fraction of such configurations which recurrs before $2N$ steps becomes small (figure \ref{s2cl}). 

It is also interesting to see that rings where $N$ is any power of 2, recurrence is seen only after exactly $2N$ time steps, irrespective of initial configuration.

\subsubsection{Quantum Kac Ring}

\begin{figure}[h!]
\centering
\begin{tabular}{cc}
\includegraphics[width=0.45\textwidth]{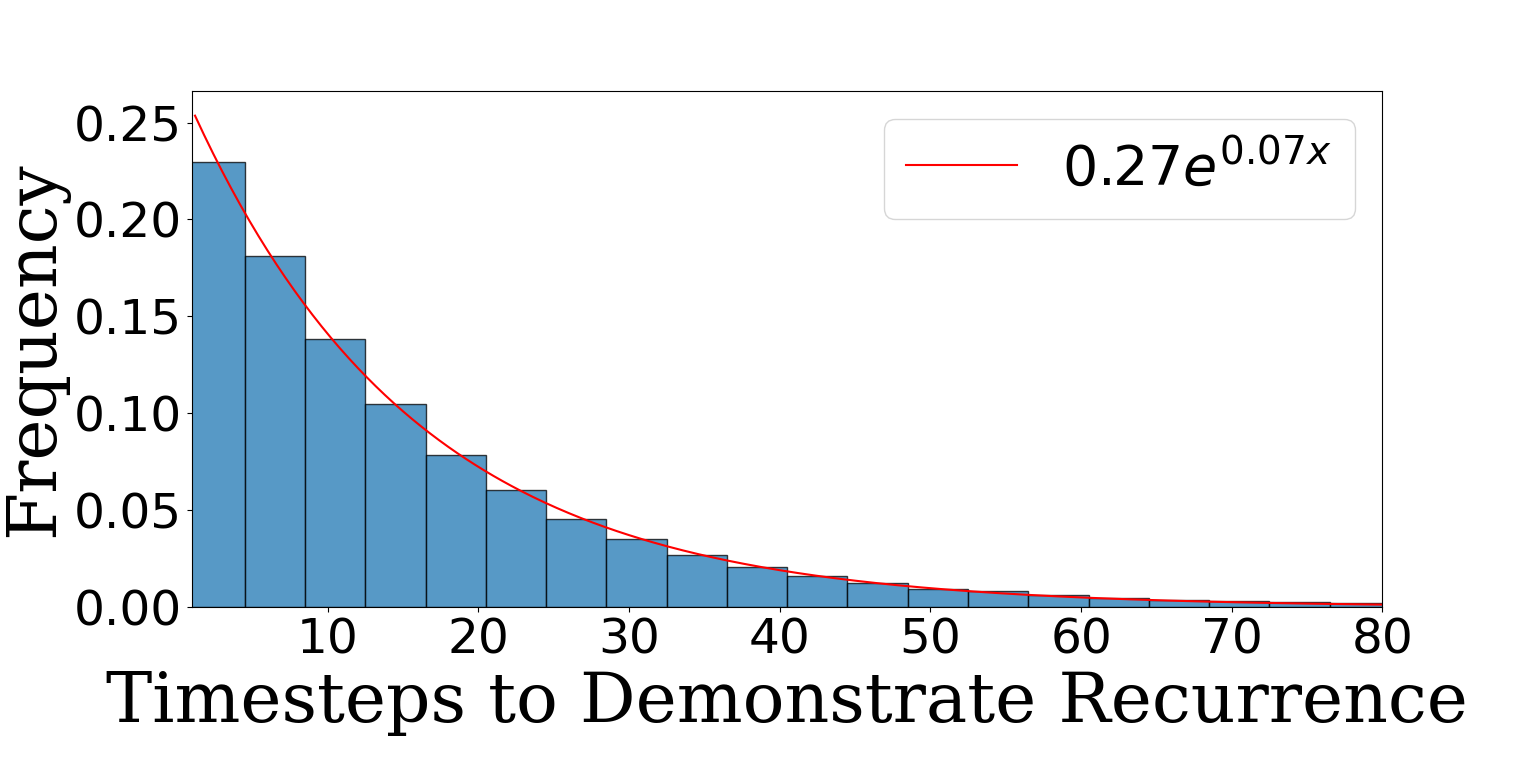} &
\includegraphics[width=0.45\textwidth]{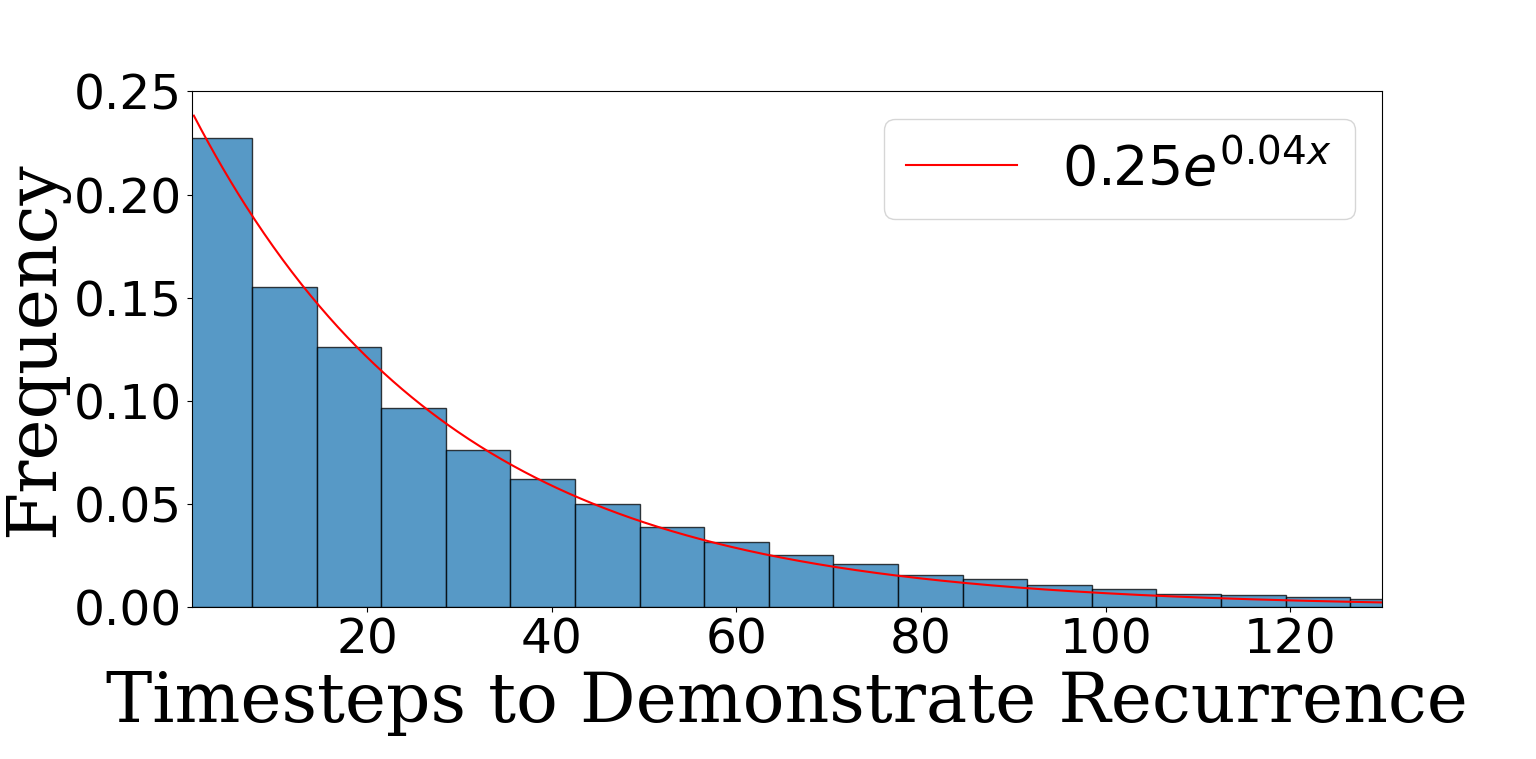} \\
(a) Quantum Ring with 4 Sites & (b) Quantum Ring with 5 Sites \\
\includegraphics[width=0.45\textwidth]{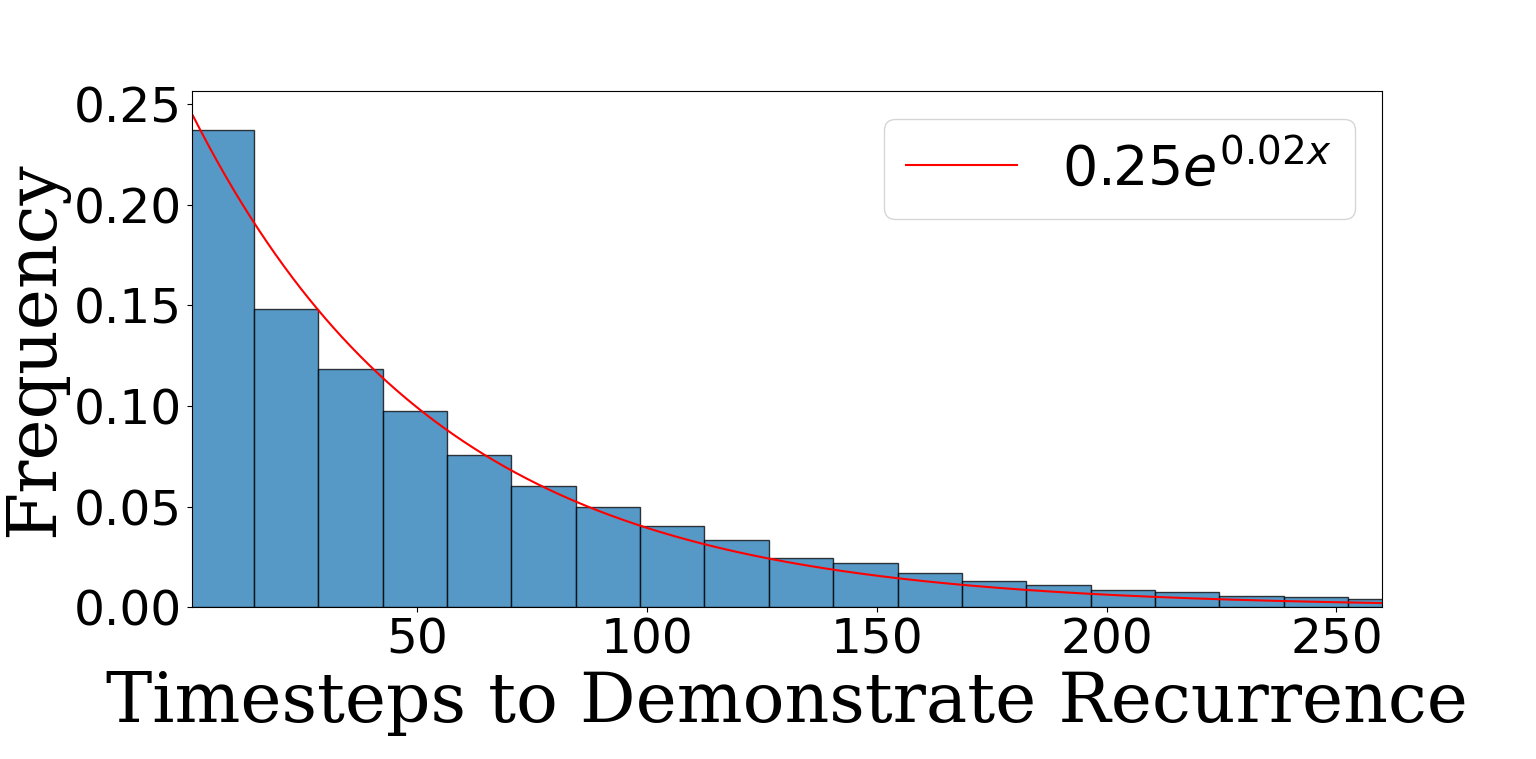} &
\includegraphics[width=0.45\textwidth]{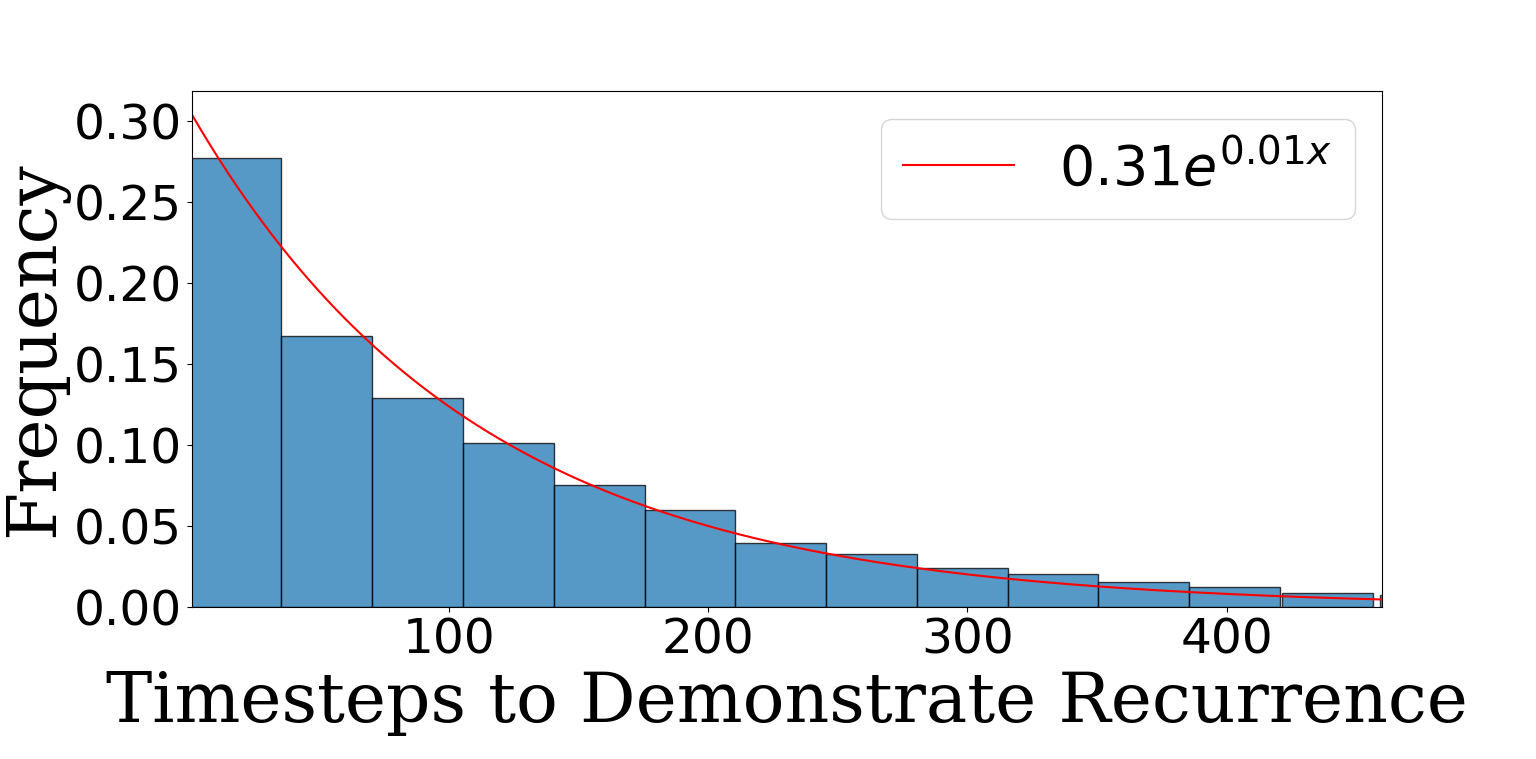} \\
(c) Quantum Ring with 6 Sites & (d) Quantum Ring with 7 Sites \\
\end{tabular}
\caption{Distribution of time steps for Poincar\'e Recurrence for a Quantum Ring}
\label{s2qt}
\end{figure}

Unlike classical Kac rings, where most time steps required for the configuration to return to its original state is $2N$, Quantum Kac rings shows a smooth variations in these values (figure \ref{s2qt}). The same initial configuration may require different time steps to return to an entropy of 0 during multiple runs depending on the results yielded by measuring the superposition of the qubit governing the pointer.

\subsection{Statistics of relative entropy}

Every run of a Kac ring follows a pattern or flow of entropy, depending on the initial configuration of the ring. Tracing these patterns for the two types of rings, classical and quantum, yields different results.
%In both, not every run has the same duration. Thus entropy trends have been stretched out to match the length of the longest run.

\subsubsection{Classical Kac Ring}

\begin{figure}[h!]
\centering
\begin{tabular}{cc}
\includegraphics[width=0.45\textwidth]{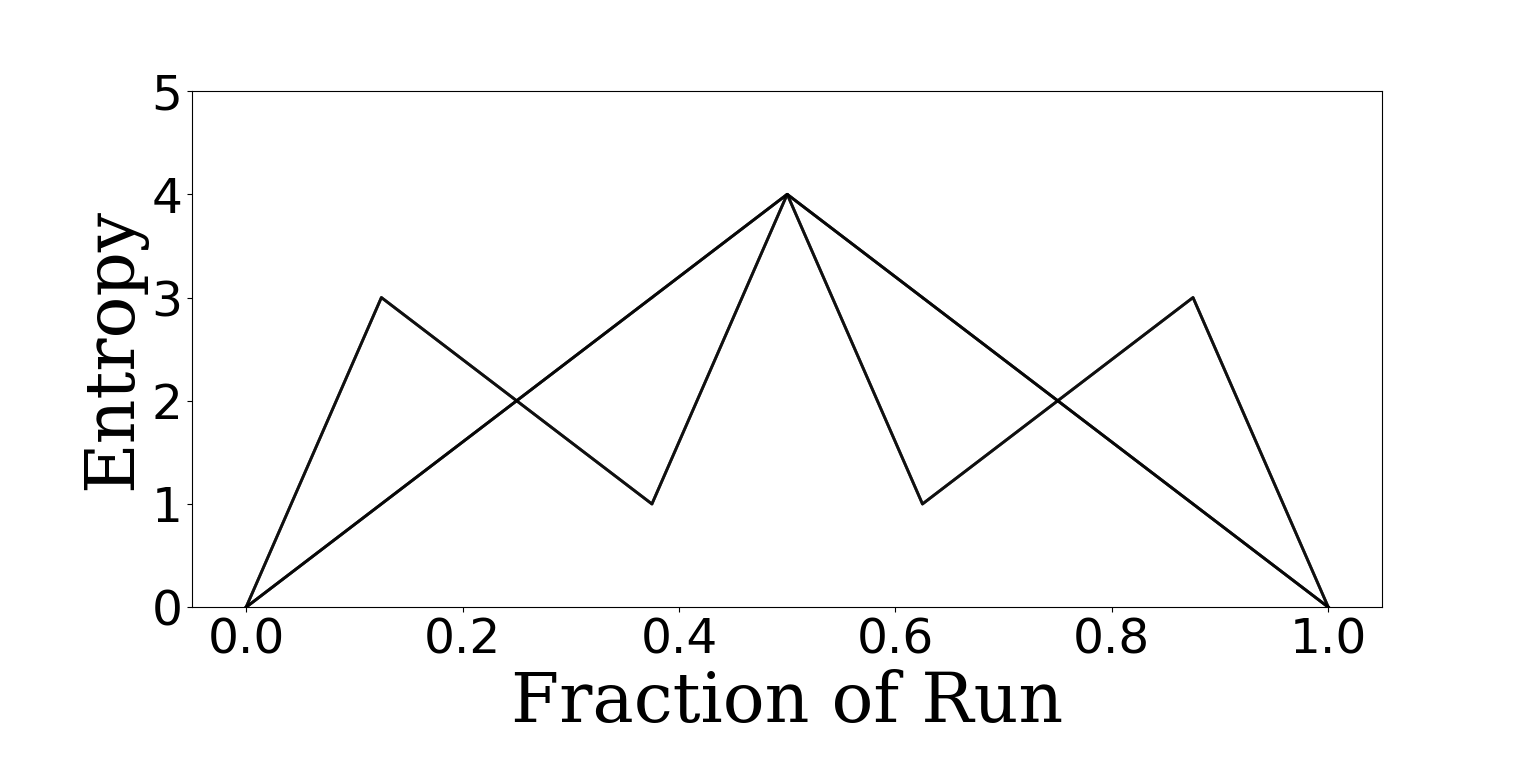} &
\includegraphics[width=0.45\textwidth]{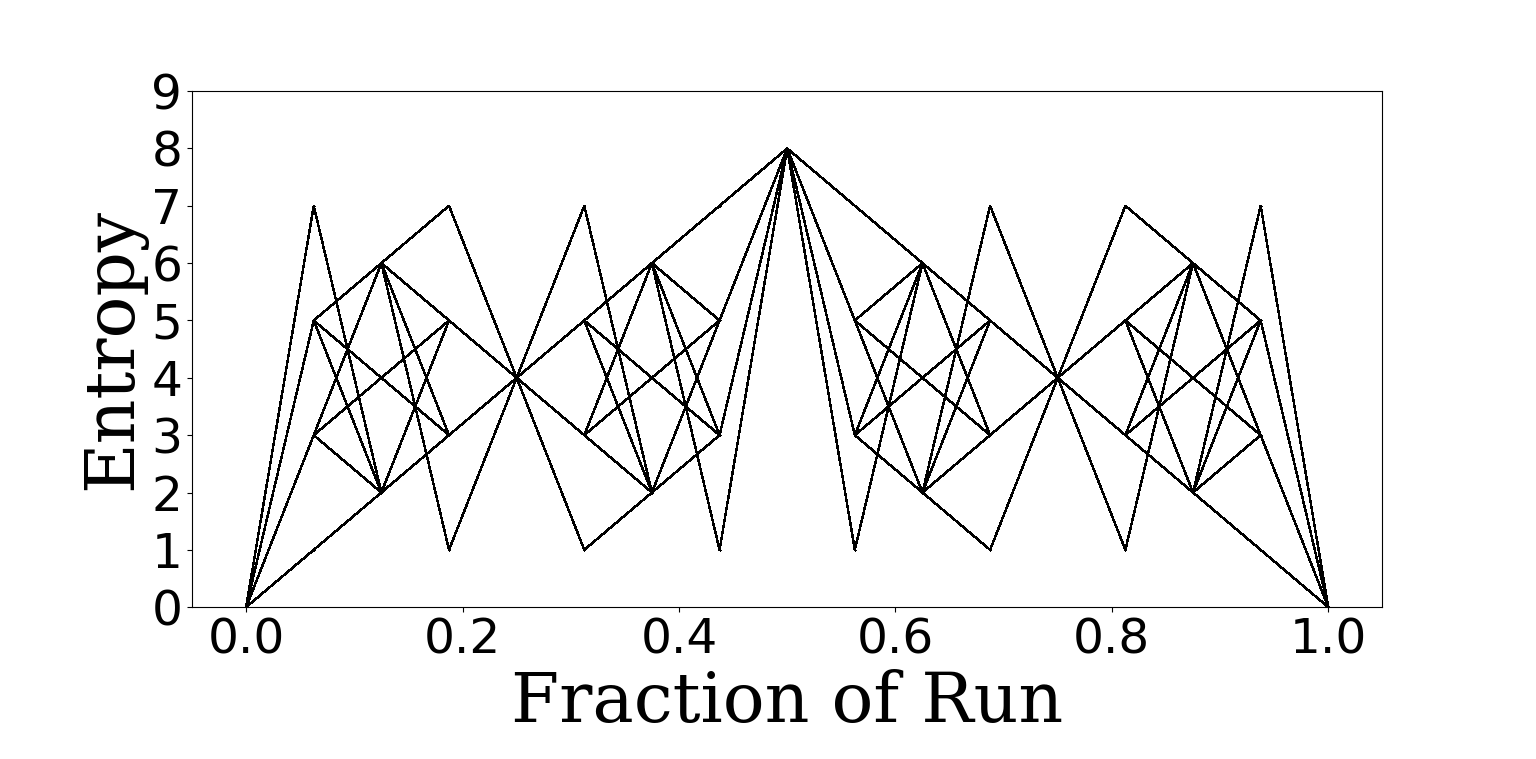} \\
(a) Classical Ring with 4 Sites & (b) Classical Ring with 8 Sites \\
\includegraphics[width=0.45\textwidth]{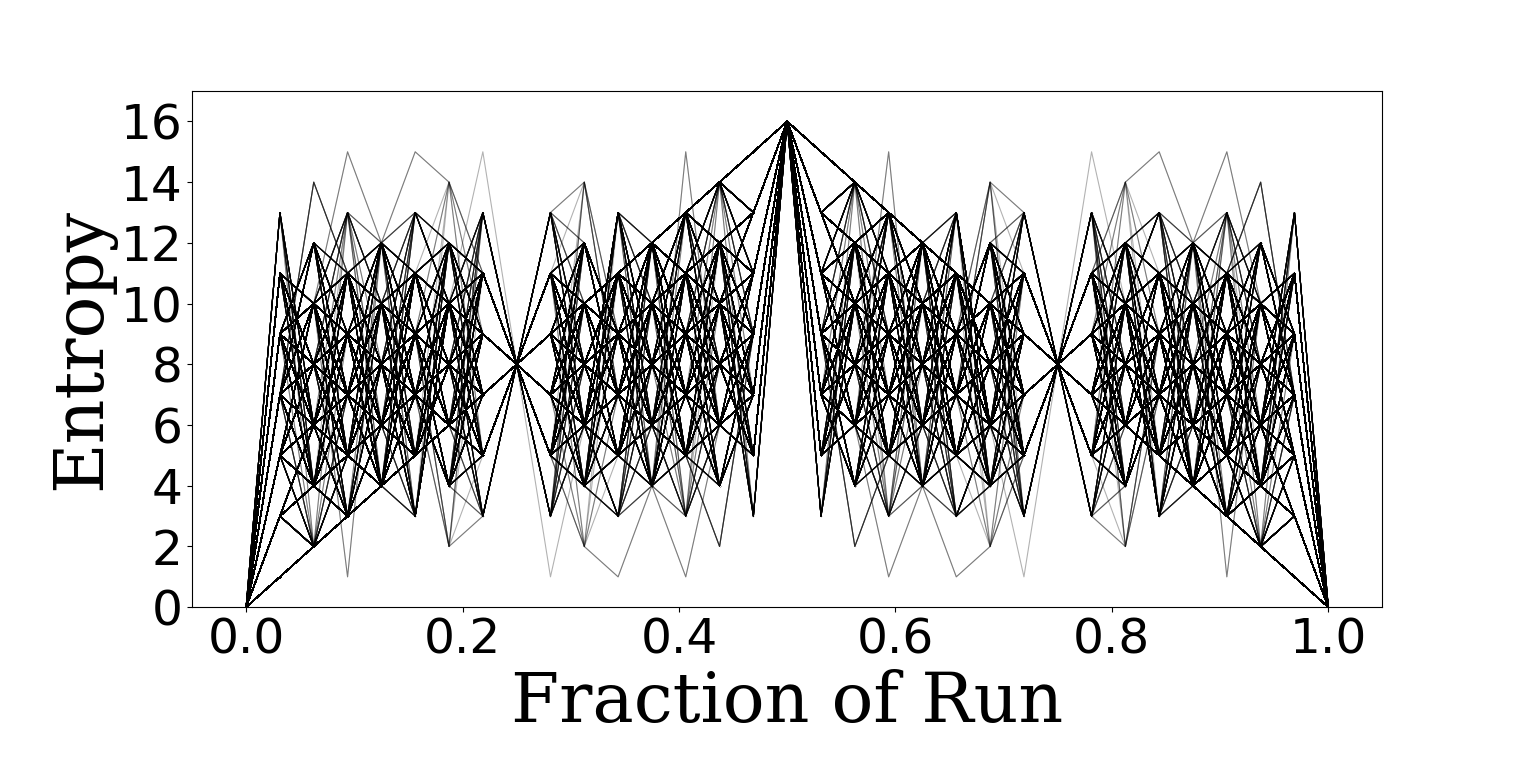} &
\includegraphics[width=0.45\textwidth]{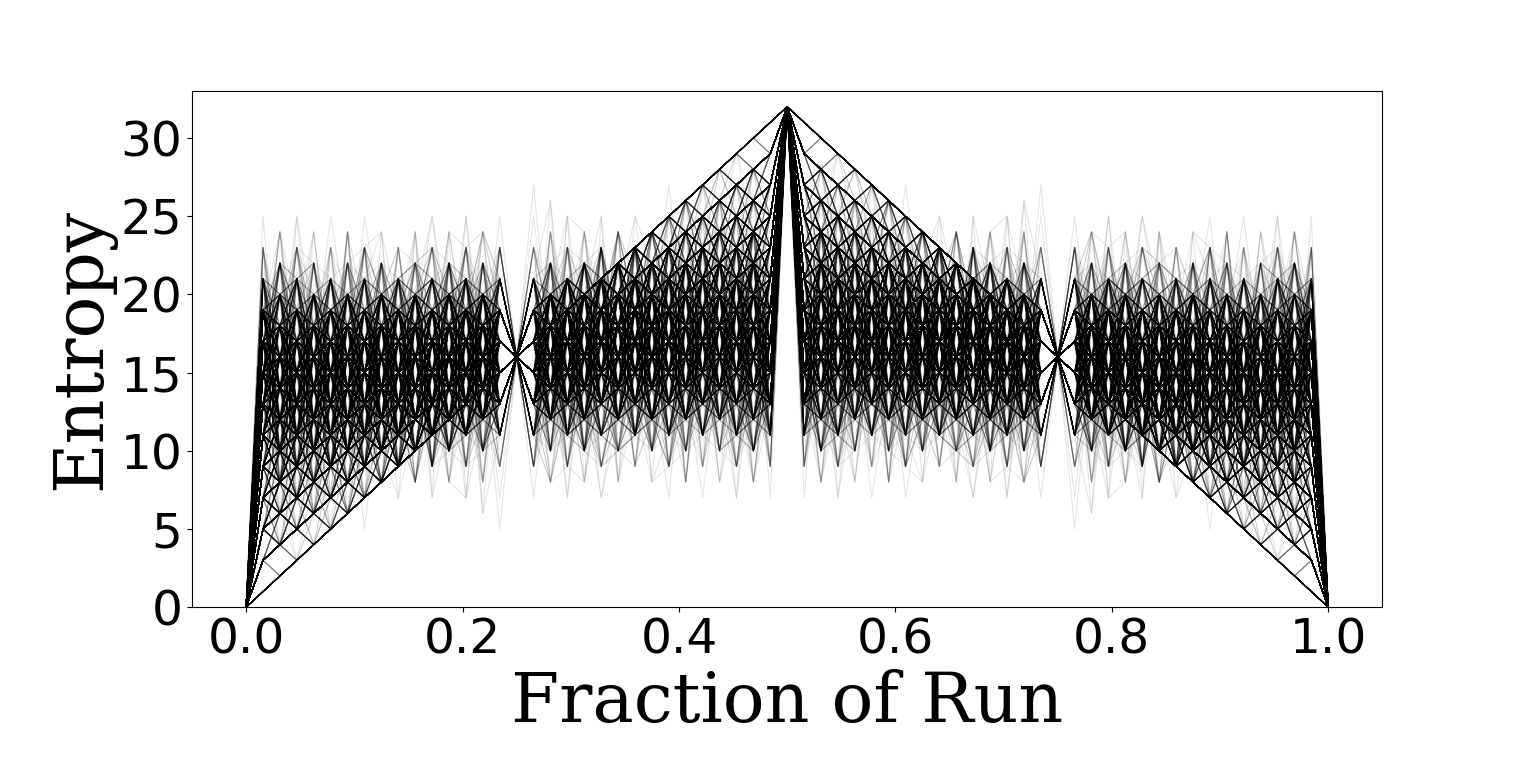} \\
(c) Classical Ring with 16 Sites & (d) Classical Ring with 32 Sites \\
\end{tabular}
\caption{Relative entropy over multiple Runs for a classical Kac ring. The time steps are normalised on the $x$-axis between 0 to 1.}
\label{s3cl}
\end{figure}

We see some interesting geometrical patterns in the case of the Classical ring (figure \ref{s3cl}). While relative entropy peaks at the $N^{th}$ timestep as expected (each site has a ball of color opposite to that of the ball there initially), the value of entropy also coincides at ($\frac{N}{2}$) and $\frac{N}{2}^{th}$ and $\frac{3N}{2}^{th}$ time steps for all the configurations. It should be noted that these are symmetric diagrams for each run and each initial configuration. The more the number of balls, the more complex these geometric diagrams become as the number of states increase.

\subsubsection{Quantum Kac Ring}
\begin{figure}[h!]
\centering
\begin{tabular}{cc}
\includegraphics[width=0.45\textwidth]{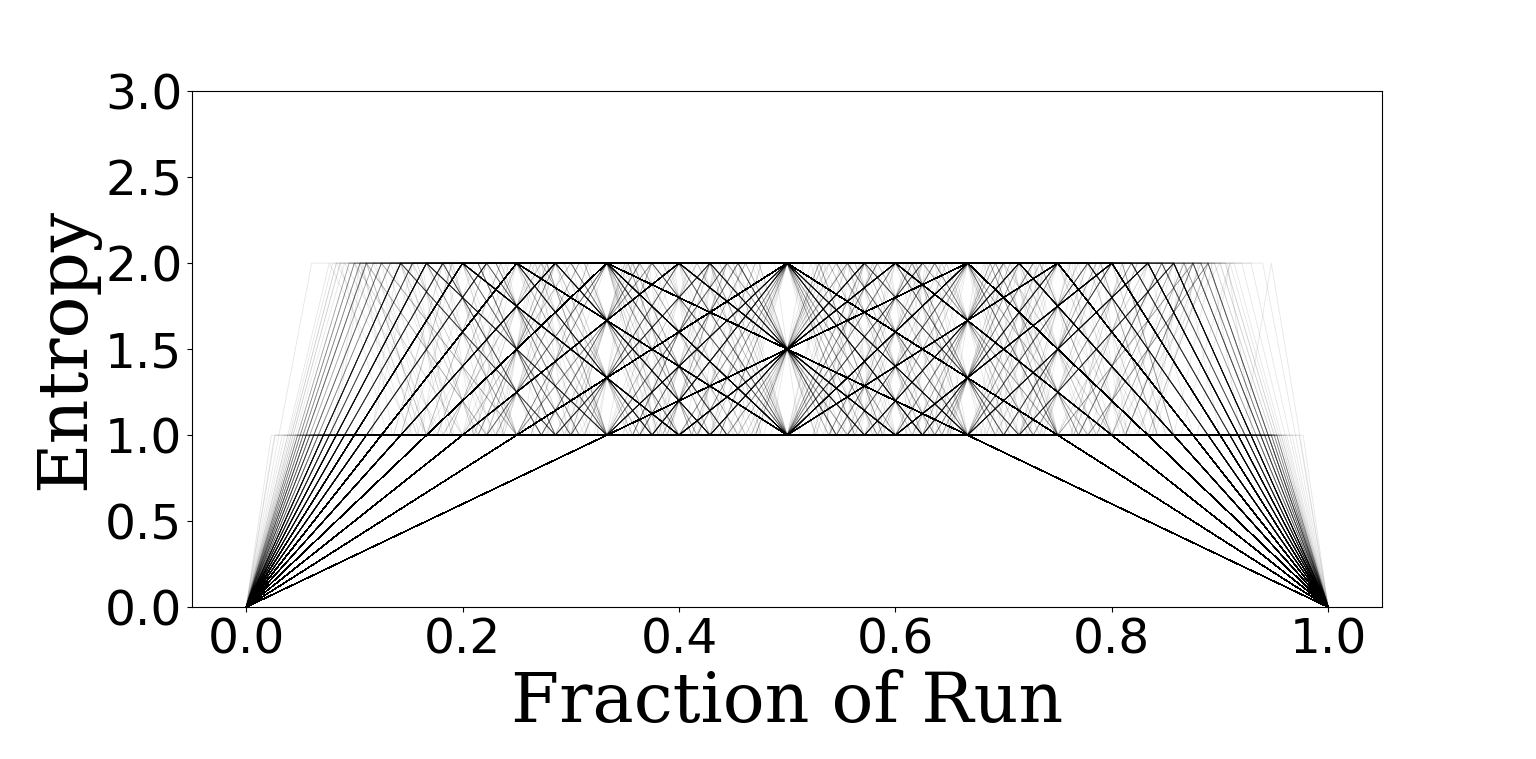} &
\includegraphics[width=0.45\textwidth]{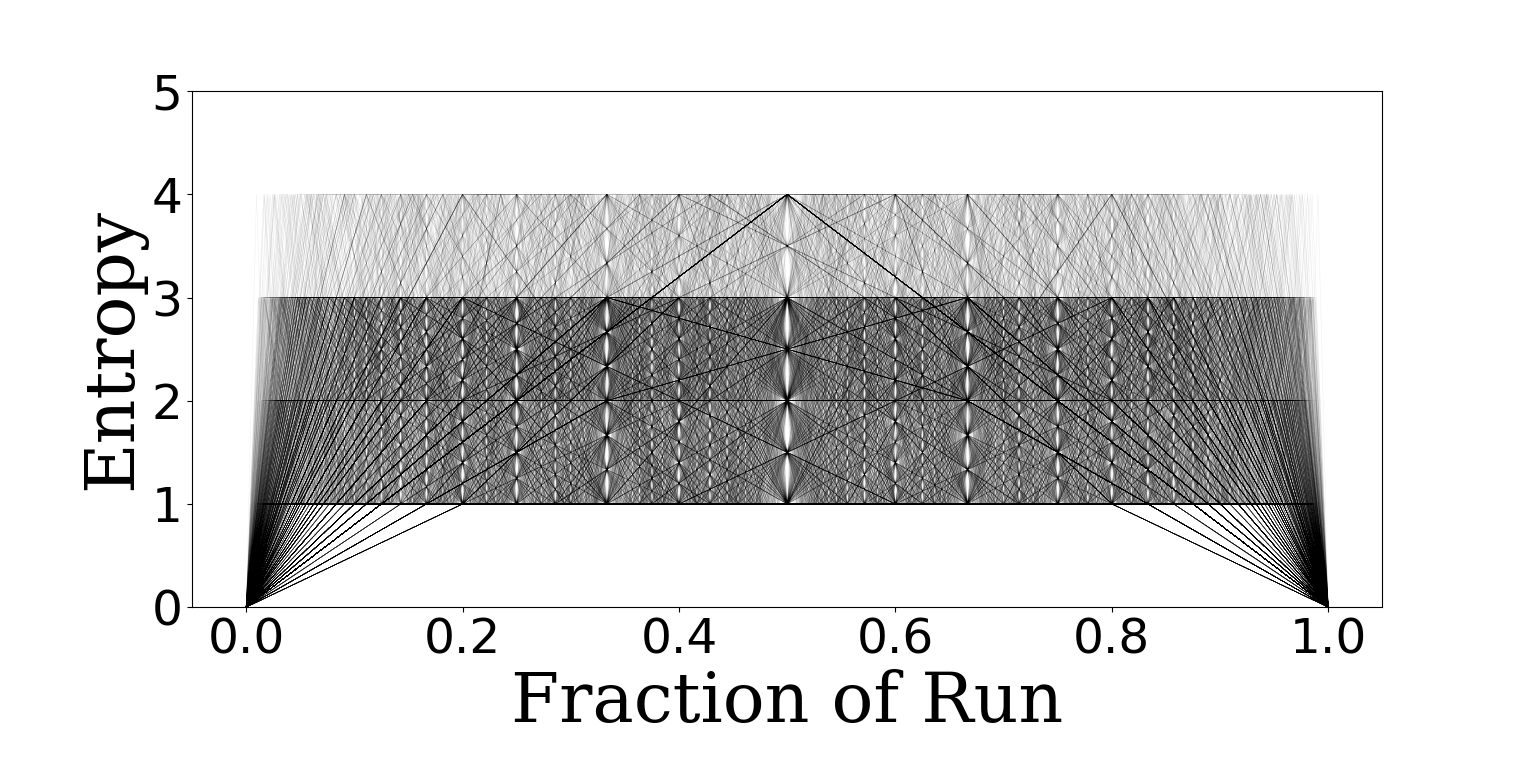} \\
(a) Quantum Ring with 2 Sites & (b) Quantum Ring with 4 Sites \\
\includegraphics[width=0.45\textwidth]{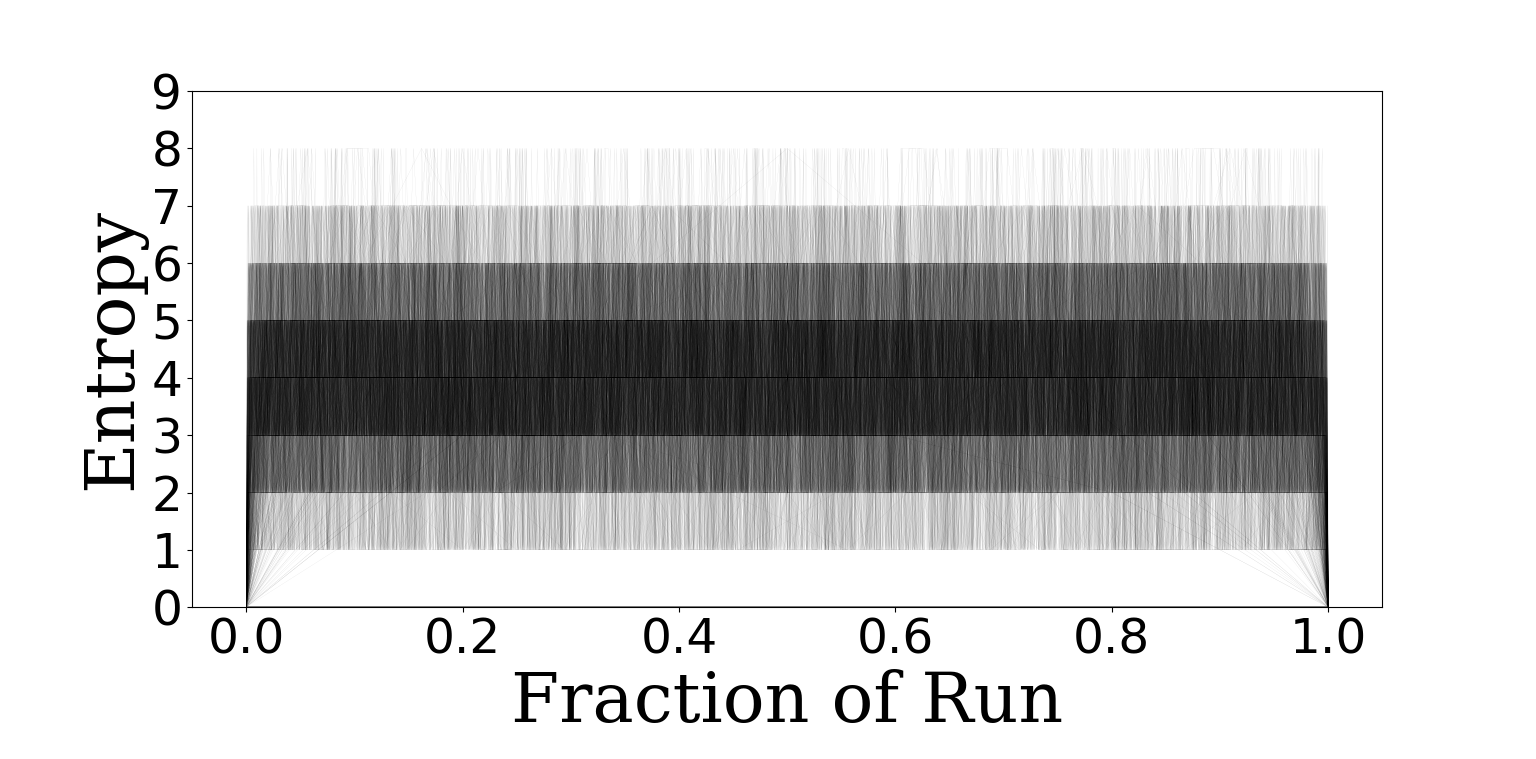} &
\includegraphics[width=0.45\textwidth]{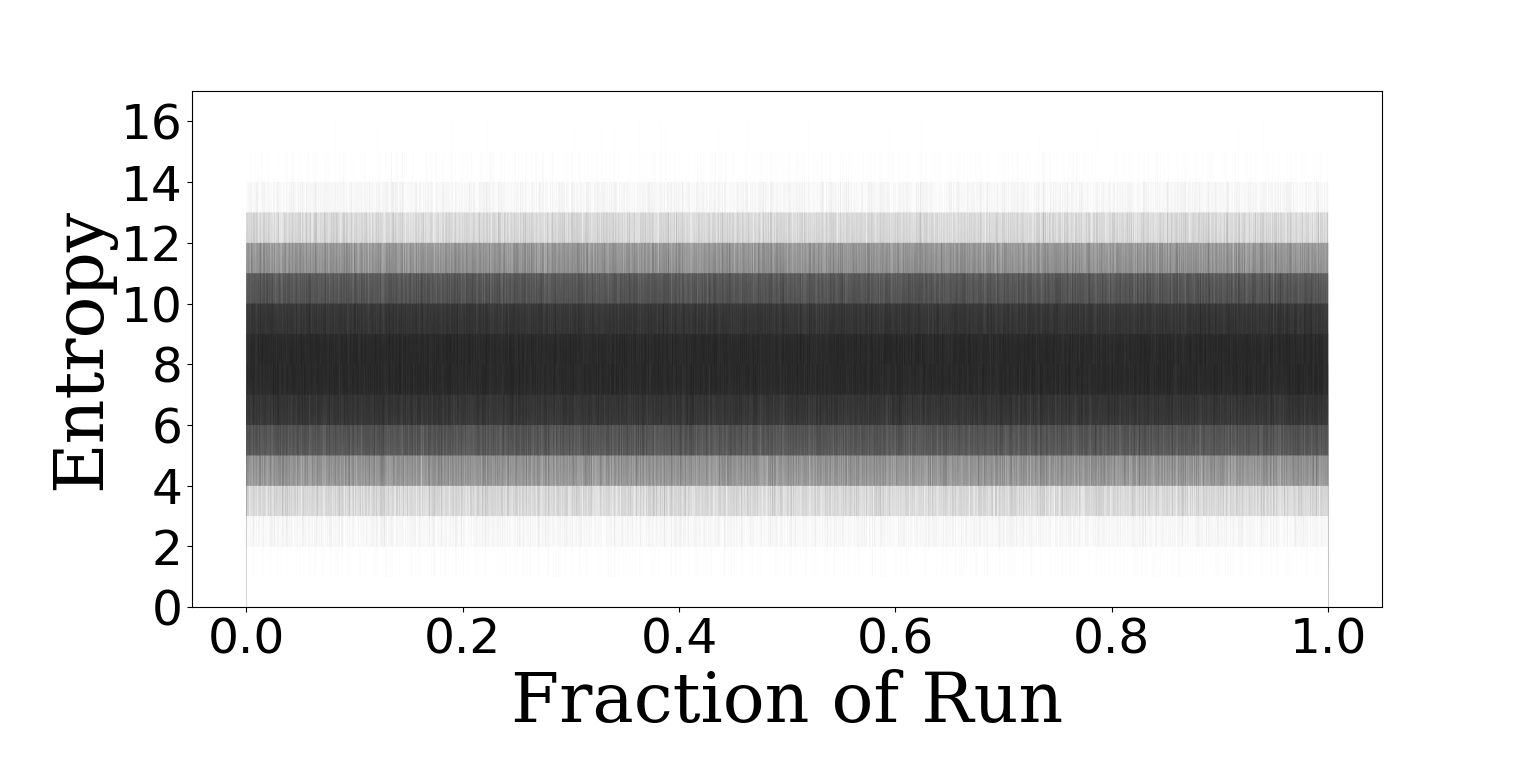} \\
(c) Quantum Ring with 8 Sites & (d) Quantum Ring with 16 Sites \\
\end{tabular}
\caption{Relative entropy over multiple Runs for a Quantum Kac ring. The time steps are normalised on the $x$-axis between 0 to 1.}
\label{s3qt}
\end{figure}

Quantum Kac rings show more intricate patterns. The classical geometric patterns wash out due to the probabilistic nature of the of the recurrence and the spread in the recurrence time (figure \ref{s3qt}). If these runs are seen individually, they may not be symmetric. The probablistic nature of the pointer paves way for many different accessible states and this leads to more intricate values of entropy to be reached.
%need to add stuff

\subsection{Time Distribution of Entropy over Multiple Runs}

Graphing the fraction of recurrence time a ring spends in a particular value of entropy also yields some interesting results. Again, recurrence time may vary over runs, especially in the case of Quantum Rings, so the time spent on each value of entropy is normalized as per the length of that run and then averaged for large ensembles.

\begin{figure}[h!]
\centering
\begin{tabular}{cc}
\includegraphics[width=0.45\textwidth]{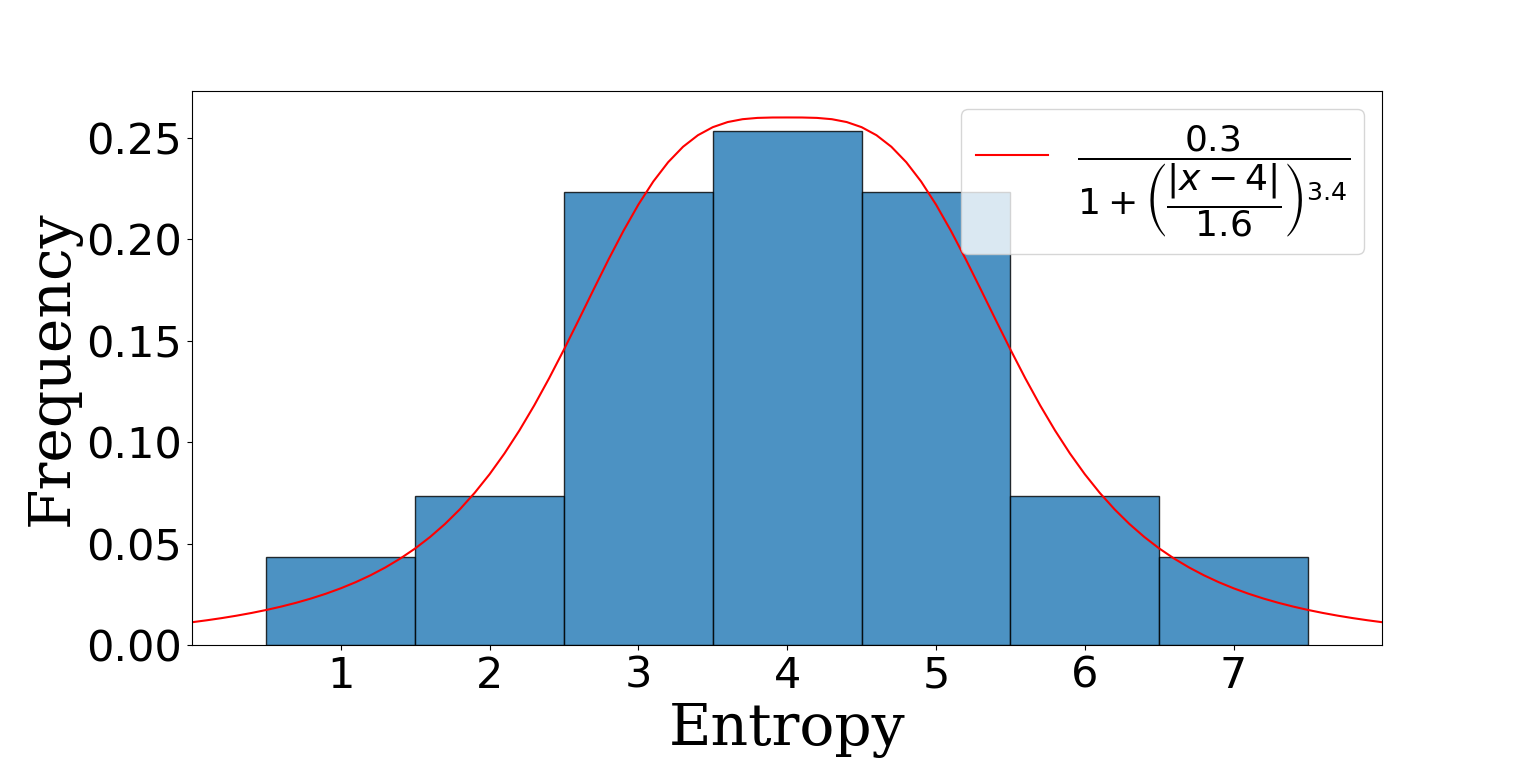} &
\includegraphics[width=0.45\textwidth]{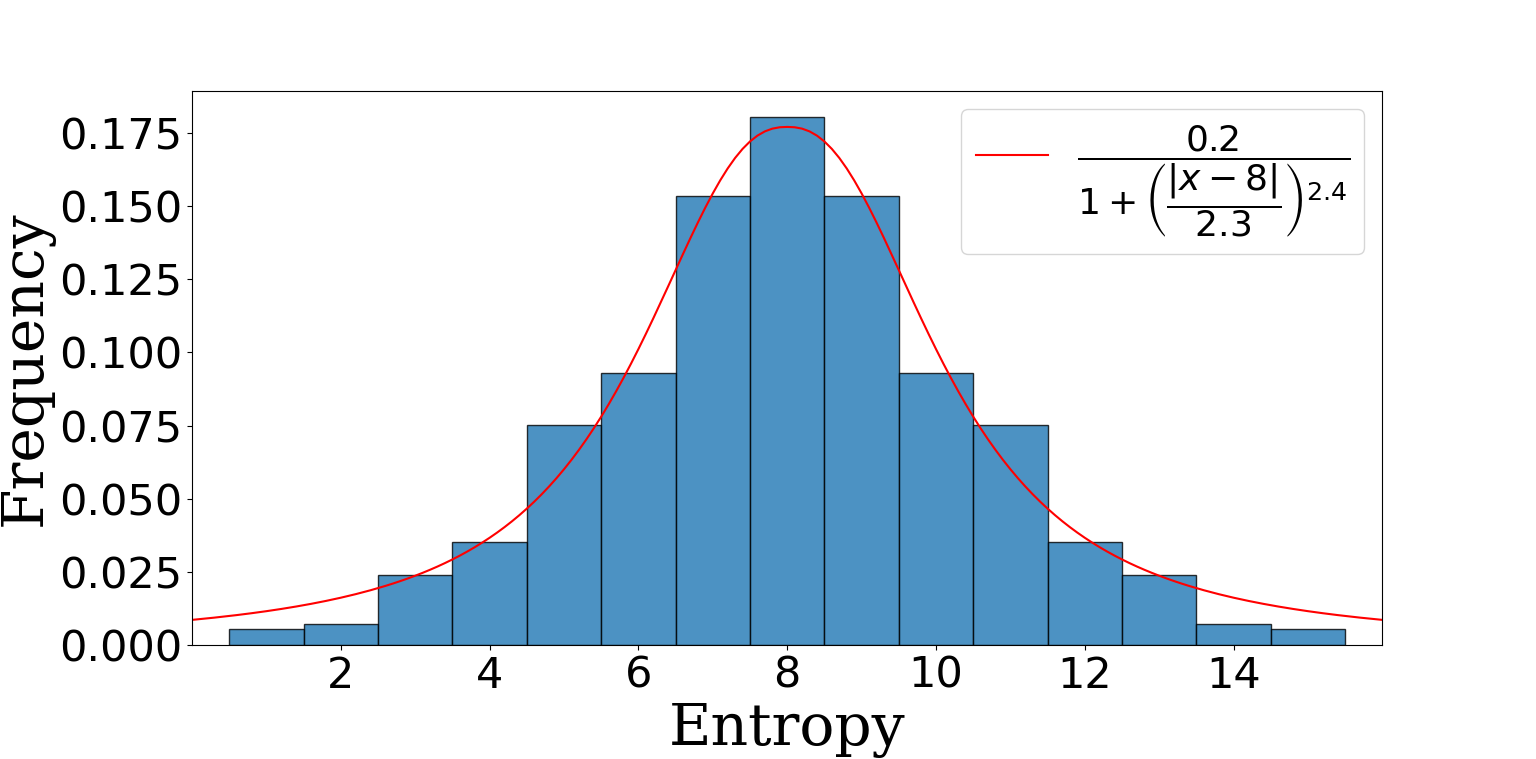} \\
(a) Classical Ring with 8 Sites & (b) Classical Ring with 16 Sites \\
\includegraphics[width=0.45\textwidth]{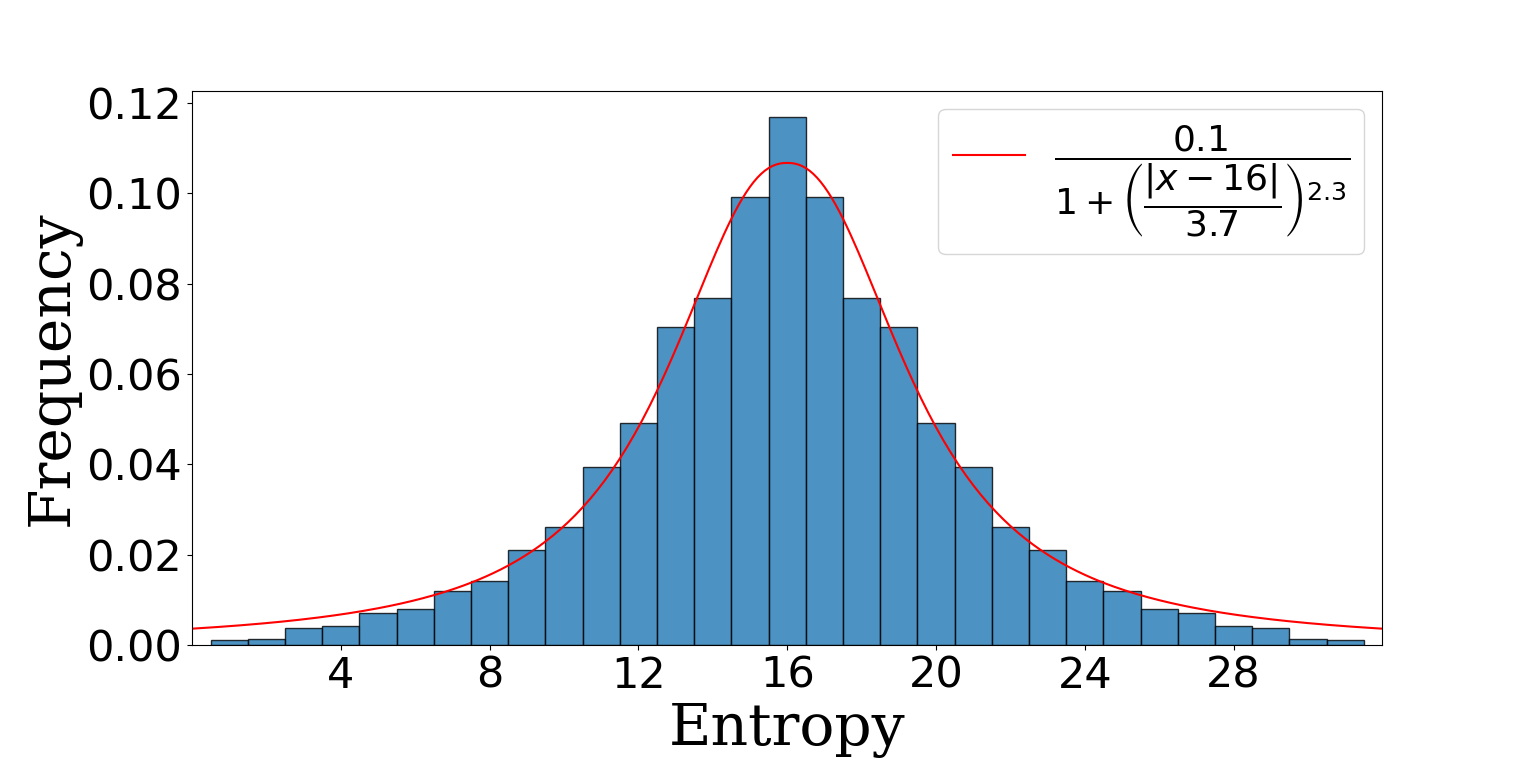} &
\includegraphics[width=0.45\textwidth]{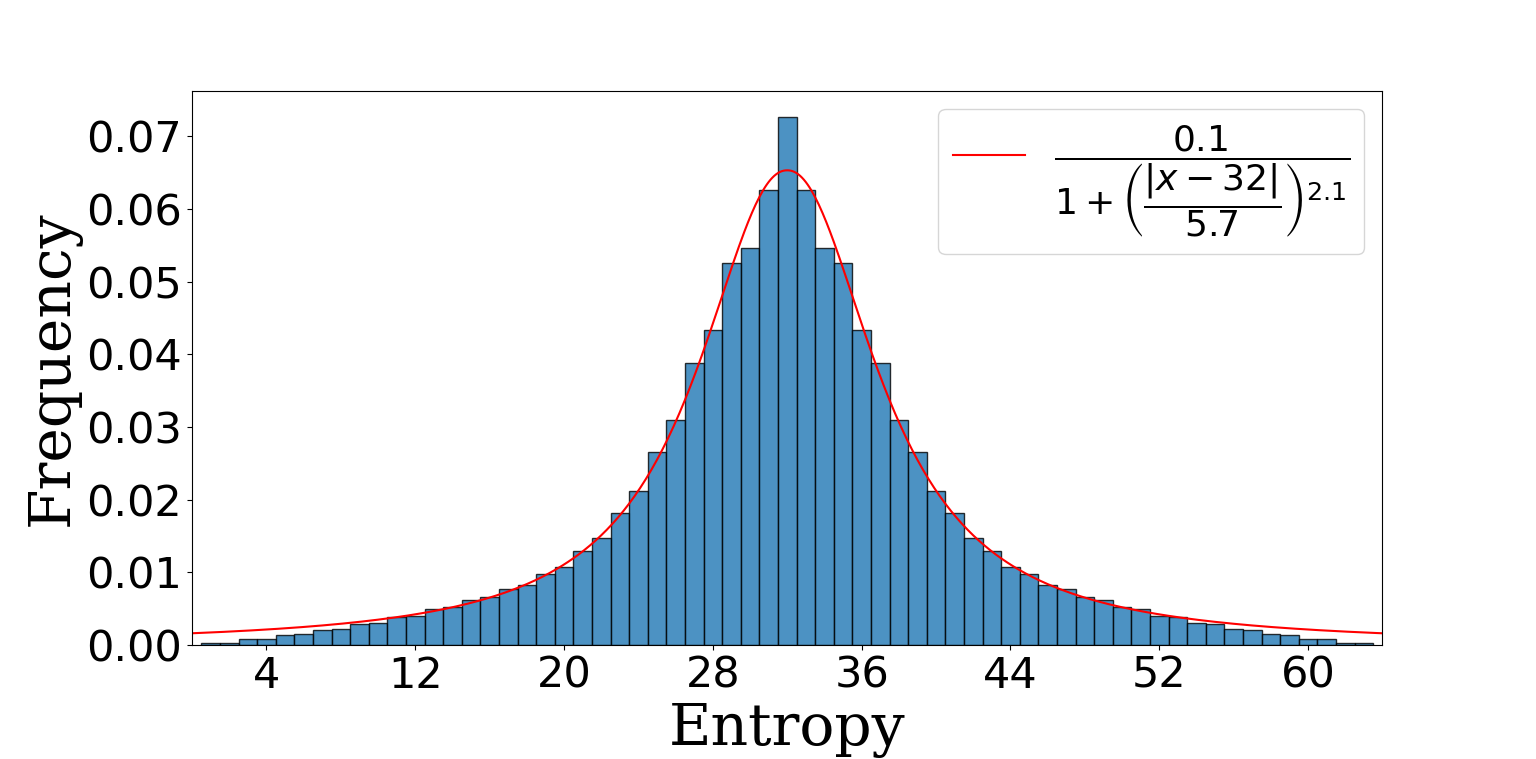} \\
(c) Classical Ring with 32 Sites & (d) Classical Ring with 64 Sites \\
\end{tabular}
\caption{Time distribution of Entropy for a Classical Kac Ring}
\label{s4cl}
\end{figure}

\begin{figure}[h!]
\centering
\begin{tabular}{cc}
\includegraphics[width=0.45\textwidth]{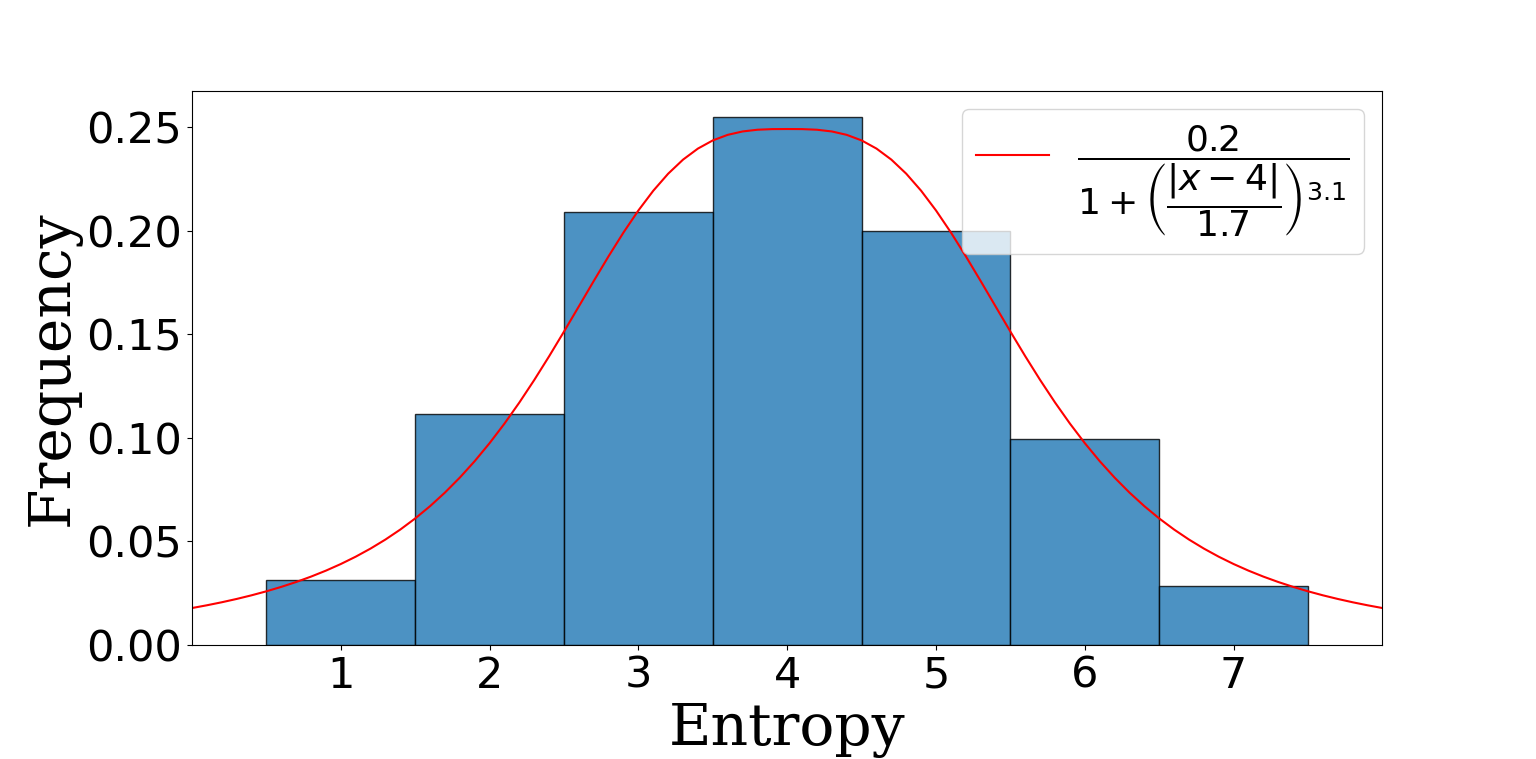} &
\includegraphics[width=0.45\textwidth]{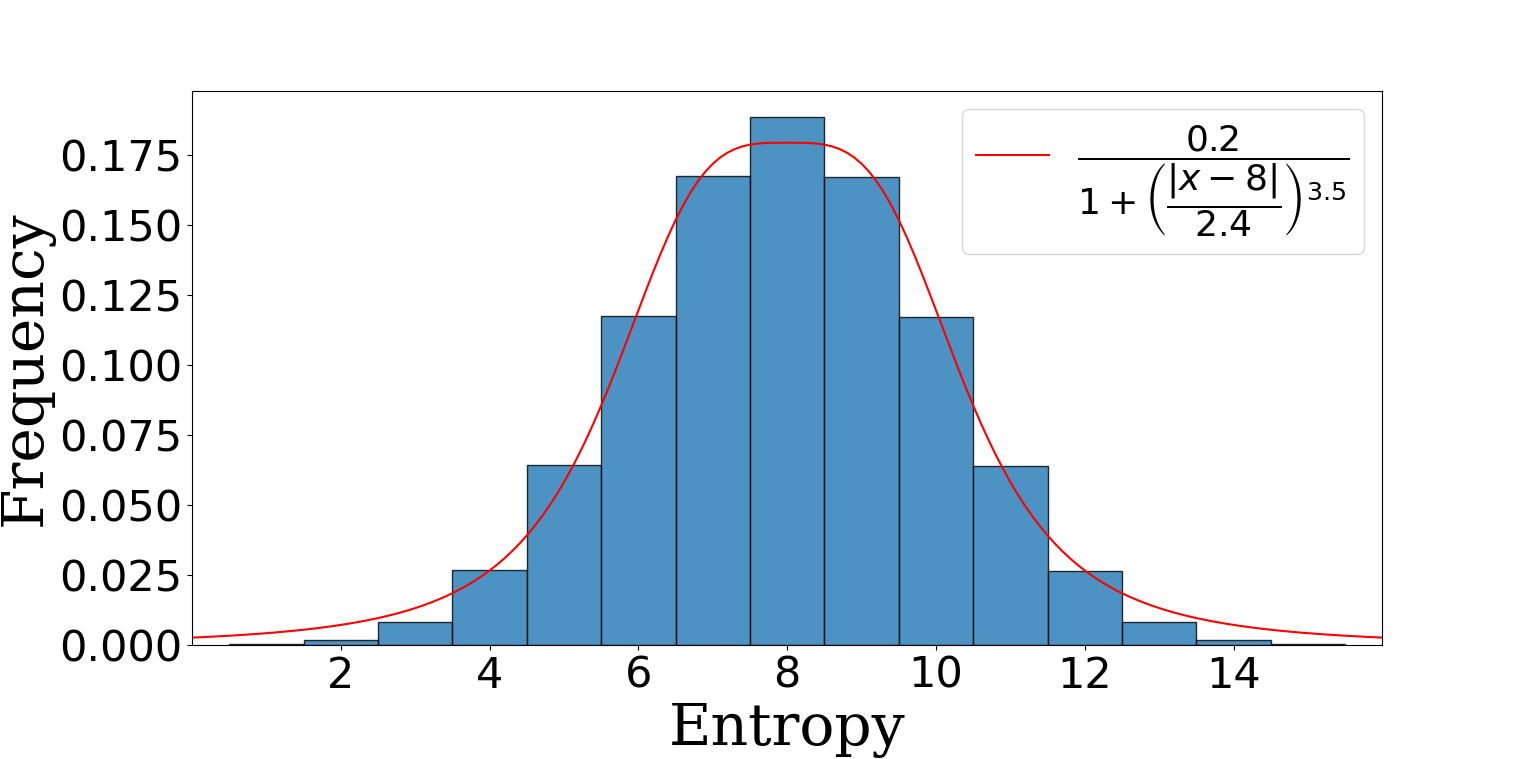} \\
(a) Quantum Ring with 8 Sites & (b) Quantum Ring with 16 Sites \\
\end{tabular}
\caption{Time distribution of Entropy for a Quantum Kac Ring}
\label{s4qt}
\end{figure}

The graph of the time distribution of the relative entropy for a Classical Kac ring (figure \ref{s4cl}) and Quantum Kac ring (figure \ref{s4qt}) is plotted .  It is compared to a general Cauchy-like probability distribution function given as 
\begin{equation}
    f(x) = \frac{a}{1+\left(\frac{x-(N/2)}{b}\right)^c}
\end{equation}
where $a,b,c$ are the fitting paramaters.

\section{Discussions}
In this work we have modeled a Kac ring with the pointer governed by the qubit. It can be thought of a stream of qubits which are regularly measured and the outcome of the measurement dictates the action of the pointer. We have simulated relative entropies and recurrence times for classical and quantum rings.

The first observation we make is that due to the pointer being governed by quantum measurements, the probabilistic behaviour skips certain balls and hence is capable for delaying the recurrence. However, certain configurations can lead to a sooner recurrence, if only certain balls are affected by the pointer. This leads to an exponential behaviour of the recurrence as we increase the number of sites and the number of balls in the ring for the quantum case, which is a linear relation for the classical case as shown in Figure \ref{s1cl} and \ref{s1qt}. This behaviour agrees perfectly to $2^N$ in Figure \ref{s1qt}.

This probabilistic nature leads to a smooth distribution in the recurrence time which can be seen in the Figure \ref{s2qt} as opposed to a very discrete one shown in the classical case (Figure \ref{s2cl}).

For a given configuration, a Quantum Kac ring can now access states of higher and lower entropies which can occur in no specific symmetric order in contrast to that of the classical case (Figure \ref{s3cl}). In the Quantum case this inclusion of probabilistic behaviour changes the distribution of entropy.

\section*{Acknowledgements}
NG and ND would like to acknowledge discussions with Rakesh Saini (Macquarie University) and Siddhant Das (LMU).

\bibliographystyle{unsrt}
\bibliography{bib}
\end{document}